\documentclass{aa}  

\usepackage{graphicx}
\usepackage{multirow,array}
\usepackage{txfonts}
\usepackage{siunitx}
\usepackage{subcaption}         % necessary for continued figures, example in section 3
\usepackage{lscape}             % to rotate a single page table, example in appendix.
\usepackage{placeins}           % useful with \FloatBarrier, to keep 
\begin{document}

   \title{A systematic assessment of the short-term X-ray behaviour of the $\gamma$ Cas analogues}

   \author{Robbie Webbe\inst{1}\fnmsep\thanks{\email{robbie.webbe@irap.omp.eu}}
        \and Yaël Nazé\inst{2}\thanks{FNRS Senior Research associate}
        \and N. A. Webb\inst{1}
        }

   \institute{IRAP, Université de Toulouse, CNRS, CNES, 9 avenue du Colonel Roche, 31028 Toulouse, France
            \and Groupe d’Astrophysique des Hautes Energies, STAR, Université de Liège, B5c, Allée du 6 Août 19c, 4000 Liège, Belgium\\}

   \date{Received September 30, 20XX}

% \abstract{}{}{}{}{}
% 5 {} token are mandatory
 
  \abstract
  % context heading (optional)
  % {} leave it empty if necessary  
   {Overall, $\gamma$ Cas analogues represent a class of OBe stars displaying bright and hard X-ray emission. This emission was recently determined to be the result of accretion onto a companion white dwarf (WD), which might possibly be a magnetic WD.}
  % aims heading (mandatory)
   {We present a systematic review of the properties of $\gamma$ Cas analogues on short-term (kilosecond, ks) scales to further investigate the nature of the X-ray source. In particular, the presence of periodicities could be associated with the rotation of a magnetic WD.} 
  % methods heading (mandatory)
   {We generated the power spectra and a generalised Fourier algorithm. We applied an epoch folding method, along with $Z^2_n$ approaches, to 131 X-ray detections of 26 $\gamma$ Cas analogues and two candidate sources.}
  % results heading (mandatory)
   {We identified five sources ($\gamma$ Cas, $\zeta$ Tau, HD45314, V771 Sgr, and $\pi$ Aqr), with recurrent periodic variability ($P\sim1.2-5.4$\,ks) across observations. Period values and pulsed fractions are consistent with those reported for intermediate polars. However, while similar periods are found in all exposures of an object, the exact period value or folded light curve shape can vary. We also identify another nine sources displaying periodicities that cannot be confirmed, as there was only one observation available for their analysis. In other cases, periodicities appear (at best) transient and/or marginally significant. Aperiodic variability could be fitted by a power law in 18 of the sources, while the power-law index always is $\sim1$, as observed in other accreting WDs.}
  % conclusions heading (optional), leave it empty if necessary
   {The global properties of the observed short-term X-ray variability of the $\gamma$ Cas analogues are consistent with those observed from other accreting WD systems. }


\titlerunning{Short-term X-ray behaviour of $\gamma$ Cas Analogues}
\authorrunning{Robbie Webbe, Yaël Nazé, and N. A. Webb}
   \maketitle
   \nolinenumbers

%%%%%%%%%%%%%%%%%%%%%%%%%%%%%%%%%%%%%%%%%%%%%%%%%%%%%%%%%%%%%%
\section{Introduction}
\label{sec:intro}

The majority of massive stars emit thermal X-rays with luminosities proportional to the bolometric luminosity following $log(L_X/L_{\text{BOL}}) \sim -7$ \citep{berghoefer_x-ray_1997,naze_hot_2009}. These X-ray emissions arise from shocks in the stellar winds and the associated plasma temperatures are rather low \citep[$T<2$\,keV, e.g.][]{waldron_extensive_2007,zhekov_x-rays_2007}. While the luminosities and/or temperatures might increase to some extent because of additional processes involving the winds (e.g. magnetically confined winds, colliding winds), a specific category of massive star has been identified by its peculiar X-ray emission: $\gamma$ Cas analogues \citep{smith_x-ray_2016}.

The $\gamma$ Cas analogues are a class of X-ray sources associated with emission-line massive stars (i.e. with spectral types OBe). Their X-ray to bolometric luminosity ratios are several orders of magnitude higher than those seen for massive stars: $\log(L_X/L_{\text{Bol}}) \sim -6.2$ to $-4$ \citep{naze_hot_2018,naze_srgerosita_2023}. While their spectra is also thermal in nature, as demonstrated by a continuum that drops at higher energies ($E>10$\,keV, \citealt{shrader_high-energy_2015}) and the presence of X-ray emission lines \citep{smith_high-resolution_2004}, they also display iron lines at 6.7 and 7\,keV, which implies unusually high plasma temperatures, exceeding 5\,keV. Other peculiarities involve large short-term variations and the presence of fluorescence lines, particularly at 6.4\,keV \citep{smith_x-ray_2016}. 

While the X-ray luminosities of the $\gamma$ Cas analogues are significantly higher than those seen in massive stars they are lower than those seen in Be-X-ray binaries (BeXRBs), where the X-ray emission is powered by accretion on to a compact object \citep{reig_bex-ray_2011,fornasini_high-mass_2023}. Nevertheless, a configuration similar to BeXRBs was envisaged for $\gamma$ Cas analogues. We note that the compact object is unlikely to be a black hole due to the relatively faint X-ray luminosities and low masses of companions. Furthermore, the presence of a neutron star has been discounted due to the lack of outbursts or of any periodic signals linked to the rotation of a neutron star typical of BeXRBs \citep[e.g.][]{reig_bex-ray_2011,fornasini_high-mass_2023}. This led to the suggestion that the emission could only be due to accretion onto a neutron star in the propeller phase \citep{postnov_propelling_2017,smith_is_2017}. However, this phase appears to be insufficiently long-lived to explain the large incidence of $\gamma$ Cas analogues among OBe stars and the link between X-ray absorption and strength of the fluorescence iron lines (as predicted for such a scenario) was not observed \citep{smith_is_2017,rauw_fluorescent_2024}.

An alternative origin theory for the X-ray emission from the $\gamma$ Cas analogues is the magnetic star-disk hypothesis \citep{smith_multiwavelength_1999,robinson_search_2000,smith_x-ray_2016}. It considers magnetic reconnection events as a result of interactions between the stellar magnetic field and magnetic fields generated in the decretion disk. These reconnection events cause the acceleration of particles towards the star, which would then lead to X-ray flares. This scenario is supported by anti-correlations between optical/UV and X-ray variability, indicating the presence of clouds near the sites of X-ray emission close to the star \citep{smith_multiwavelength_1998}, and the lack of a delayed correlation between optical and X-ray variability, which would indicate transit times following ejection of material from the star before accretion onto a companion \citep{motch_origin_2015}. However, recent XRISM data revealed that the signatures of the hot X-ray plasma follow the motion of the companion and not the Be star \citep{naze_xrism}, leading to the rejection of this scenario.

A final model employed to explain the peculiar X-rays from $\gamma$ Cas analogues considers accretion onto a white dwarf (WD, \citealt{murakami_x-ray_1986}). This model is compatible with the detected plasma motion as well as with the binarity properties of a number of the $\gamma$ Cas analogues \citep{harmanec_properties_2000,nemravova_properties_2012,bjorkman_study_2002,naze_velocity_2022,naze_x-raying_2024}. Indeed, in these binaries, the companion has gone undetected thus far; for example, interferometric campaigns have resulted in non-detections of stripped He-star companions in the observed $\gamma$ Cas analogues \citep{klement_chara_2024}. In addition, the probable masses of the companions are compatible with WD mass ranges \citep[$M\sim1M_{\odot}$: e.g.][]{nemravova_properties_2012,naze_velocity_2022,naze_longterm_2026}. Furthermore, the strengths and ratios of the iron lines at 6.4--7.0\,keV appear similar to those seen in identified cataclysmic variables (CVs) and symbiotic stars \citep{mukai_x-ray_2017}. In this context, it is important to note that the confirmed orbital periods of these systems are typically of the order of hundreds of days, as expected from population synthesis models for Be binaries \citep{shao_population_2021}. 

The $\gamma$ Cas analogues display significant X-ray variability on a wide range of timescales, ranging from seconds to years. On the shortest timescales, below hundreds of seconds, the X-ray variability of the $\gamma$ Cas analogues appears to be consistent with a stochastic 'flicker' noise. This short-term variability (i.e. shots \citealt{smith_x-ray_2016}) is well-described as a series of flares, and power spectral analysis of emission on these timescales in several $\gamma$ Cas analogues is approximately consistent with a $1/f$ power spectrum \citep[e.g.][]{parmar_x-ray_1993,lopes_de_oliveira__2010,torrejon_chandra_2012,naze_x-raying_2024}. This is consistent with that observed in some CVs \citep{naze_x-raying_2024, gunderson_time-dependent_2025}. Additional short-term variations are limited to the softer X-rays ($E\leq1$\,keV). They are due to (local) absorbers appearing in the line of sight \citep{hamaguchi_discovery_2016,smi19,gunderson_time-dependent_2025}. 

Between observations, on timescales from several days to multiple years, variability in the X-ray luminosity is often detected and can reach one order of magnitude or more. The causes of this long-term variability are not clear, but changes in the broadband optical brightness have been found to correlate with X-ray flux in several $\gamma$ Cas analogues \citep{motch_origin_2015,rauw_intriguing_2018,naze_x-ray_2022, naze_longterm_2026}. There has been no link detected, however, between the flux changes and the orbital phase or the strength of H$\alpha$ emission \citep[e.g.][]{naze_surprises_2019,rauw_x-ray_2022,naze_x-raying_2024}. Ultimately, the understanding of these long-term variations is constrained by the limits of monitoring and observation of long-term trends over several years simultaneously in optical and higher energy ranges.

On intermediate timescales (i.e. kilosecond scales), there have been several claimed detections of potential periodicities for a number of $\gamma$ Cas analogues. In the archetype $\gamma$ Cas, the first claimed periodicity came nearly four decades ago from an analysis of an \emph{EXOSAT} observation, where an epoch folding search identified a periodicity of 6\,ks \citep{frontera_time_1987}. This could not be confirmed in later \emph{EXOSAT} observations \citep{horaguchi_be_1994}. Later \emph{ROSAT} observations were identified as possibly containing a periodicity of 8.1\,ks \citep{haberl__1995}. More recently, a significant periodicity of 3.4\,ks was identified in one observation among a sequence of six \emph{Chandra} HETG observations of $\pi$ Aqr in 2022 \citep{huenemoerder_chandra_2024} using an epoch-folding period search. In parallel, a search for periodicities using the $Z^2_n$ statistic identified a potential periodicity of 3.2\,ks, with a pulse fraction of 24 \%, in a 2004 observation of HD\,161103 \citep{lopes_de_oliveira_new_2006}.  In addition, a periodicity of 3.7\,ks was reported by \citet{naze_x-raying_2024} in one \emph{XMM-Newton} observation of $\zeta$\,Tau following a Fourier analysis but there was no identified periodicity in the second observation considered in this analysis. Finally, a 4.1\,ks periodicity was also detected in V771\,Sgr by \citet{mondal_periodicity_2024} during a blind search for periodic X-ray sources in an \emph{XMM-Newton} survey of the Galactic Plane. An analysis of the same observation by \citet{webbe_stonks_2026} using epoch folding and $Z^2_n$ statistics identified a significant periodicity at nearly half of this value (i.e. 2.08\,ks). 

The significance of these detections varies greatly, and these periodicities have not yet been robustly identified as persistent across multiple observations. However, a recurrent detection would be an important indicator for the nature of $\gamma$ Cas analogues. Indeed, such a variability would point towards a periodic phenomenon; in particular, the rotation period of a WD companion, as observed in magnetic CVs \citep{kuulkers_x-rays_2006,mukai_x-ray_2017,webb_accreting_2023}. 

In this paper, we analyse all available \emph{XMM-Newton} and \emph{Chandra} observations of $\gamma$ Cas analogues and detail our systematic review of the temporal variability on kilosecond timescales. In Section \ref{sec:methods}, we describe the methodology applied to these X-ray observations for the characterisation of temporal variability and the detection of significant periodicities. In Section \ref{sec:freq_dep_var}, we present and discuss frequency-dependent aperiodic variability trends. In Section \ref{sec:period}, we identify any significant periodicities and put these results in the context of the models for X-ray emission from the $\gamma$ Cas analogues. In Section \ref{sec:conc}, we present the conclusions of this systematic review of periodicities in the $\gamma$ Cas analogues.
\vspace{-0.2cm}

\section{Data and methods}
\label{sec:methods}

In this analysis, we review X-ray observations from the \emph{XMM-Newton} and \emph{Chandra} observatories to determine the presence or absence of significant temporal periodicities on kilosecond scales. This list of sources is motivated by the populations of $\gamma$ Cas analogues, as curated by \citet{smith_x-ray_2016}, \citet{naze_let_2020}, \citet{naze_three_2020}, and \citet{naze_x-raying_2024}.

\subsection{Observational data}
\label{subsec:obs_data}

We considered all pointed \emph{XMM-Newton} and \emph{Chandra} observations with detections of the 26 $\gamma$ Cas analogues and 2 candidates (see Appendix \ref{app:obs-list}). These observations vary in length from 2.2\,ks to 141.3\,ks, with an average exposure of $\bar{T} = 24.3,ks$ and were largely free from flaring and other observational effects that could have affected the observing time.

The \emph{XMM-Newton} observations were reprocessed with XMMSAS version 21.0 and \emph{Chandra} Level 2 products as per \texttt{ciao v4.16}. Light curves and event lists for all sources were extracted on an observation-by-observation basis with source and background regions manually selected to provide high signal-to-noise ratio (S/N) for the source while avoiding nearby sources. Photon events were then barycentre corrected using the \texttt{barycen} task for \emph{XMM-Newton} observations and the \texttt{axbary} task for \emph{Chandra} observations. For all the observations, we then created the light curves with time bins of 10\,s for use in the time- and Fourier-domain analyses. Due to the high observed count rate in \emph{XMM-Newton} observations of $\gamma$ Cas, and the subsequent risk of pileup, we only used the photon event data from the EPIC pn instrument. Indeed, the pn instrument was in timing mode for observations 0201220101, 0840310101, 0840310201, 0840310301, and 0840310401, enabling the observation of X-ray bright sources. For the other five observations, we checked that the EPIC pn observations were free from pileup using the XMMSAS task \texttt{epatplot}.

\subsection{Fourier analysis}
\label{subsec:meth_fourier}

All observations were examined for periodicity in the frequency-domain. We employed two approaches to mitigate the effects of fragmented observations affected by background flaring. First, for individual good time intervals (GTIs) within the observations, the sampling was regular; thus, we took the discrete Fourier transform of the light curve within this range and squared the complex result to obtain the power spectral density (PSD). To calculate the frequency-domain products we used light curves with time bins of 10\,s in the energy range 0.2-12.0\,keV. For observations with several GTIs, we created PSDs for any GTIs in light curves that were longer than 10\,ks or the single longest GTI in the event where none of them exceeded 10\,ks in length.

In our work, the PSDs were calculated using the \texttt{PowerSpectrum} class in the \texttt{stingray} \citep{huppenkothen_stingray_2019} module for \texttt{Python}. The PSDs were manually examined to determine frequency ranges that were not dominated by Poisson noise, which were then fitted using a power-law relationship of $PSD(f) \propto f^{-\alpha}$. Next, the significances of any observed peaks in the PSDs in these frequency ranges were determined using the approach of \citet{vaughan_simple_2005}, as a comparison against random peaks that could appear in a PSD dominated by (red) noise (see also Eq. \ref{eq:psd-sig}). As an initial selection, we only considered peaks in the PSDs with significance of $3\sigma$ or greater, with a lower limit on the frequency of $4/T$, where $T$ is the length of the observation and frequency steps of $\Delta f = 1/T$. This lower limit is chosen to ensure that any observed periodicities contain at least four cycles in a given observation. The upper limit is chosen on an observation-by-observation basis to select sections of the PSD that are not dominated by Poisson noise. Thus, we determined the significance of any peaks as
\begin{equation}
\label{eq:psd-sig}
      \gamma_{\epsilon} = -2ln[1-(1-\epsilon)^{1/n'}],
\end{equation}
where $\epsilon$ is the required significance probability ($\epsilon=0.0013$ representing a 3$\sigma$ significance) and $n'$ is the number of points from the PSD under consideration. The power-law fit to the PSD is multiplied by $\gamma_{\epsilon}/2$ to determine the threshold for significance at each frequency.

Second, we used a generalised Fourier algorithm (GFA) adapted to uneven samples \citep{heck_period_1985,zechmeister_generalised_2009} for deriving the frequency spectra associated to entire observations as they could be affected by uneven sampling because of gaps linked to bad time intervals. GFAs were calculated for frequencies in the range $ 4/T \leq f < 0.05$\,Hz (the Nyquist frequency for 10s-binned light curves) in the case of \emph{XMM-Newton} EPIC pn observations and with an upper limit of 0.01\,Hz in the case of \emph{XMM-Newton} EPIC MOS and \emph{Chandra} observations. We set a more stringent upper frequency limit for \emph{XMM-Newton} EPIC MOS and \emph{Chandra} observations due to the longer frame times for these instruments compared to that of the \emph{XMM-Newton} EPIC pn camera to prevent the detection of spurious periodicities at multiples of the instrumental frame time. The frequencies were oversampled by a factor of 25 (frequency steps $\Delta f = 1/25T$) to allow us to probe a finer range of frequencies in this interval. We fit the low-frequency continuum variability of the GFA to assess the observed aperiodic variability (squared amplitudes $A(f)^2\propto f^{-\alpha}$, as above), but the approach of \citet{vaughan_simple_2005} cannot be applied to assess the peaks in the GFA as it requires evenly sampled data. We therefore determined the significance of any peaks in the GFA for each observation using the approach described by \citet{baluev_assessing_2008}. This approach calculates an analytical upper limit on the false alarm probability using extreme value statistics, and is comparable in accuracy to the use of Monte-Carlo simulations that would not be feasible for this large review of observations. The GFAs were created using the \texttt{LombScargle} class implemented in the \texttt{astropy} module for \texttt{Python}, and the significance of any peaks in the GFAs were calculated using the Baluev approach implemented under the \texttt{false\_alarm\_probability} method in \texttt{astropy}. As before, we consider a peak significant if it reaches the one-sided 3$\sigma$ level or a Baluev false alarm probability of $p=0.0013$. This test is formally valid in the presence of white noise, while we expect some red noise (see above). However, the results demonstrate that this caused no problems, with the detected frequencies showing a consistency with those found from other methods.

\subsection{Photon event folding}
\label{subsec:meth_pe_folding}

We also determined the presence of periodicities in the X-ray emission from the $\gamma$ Cas analogues by analysing observations in the time-domain. This was achieved by folding the arrival times of photons over a range of periods, and determining the significance of variations in these folded light curves using the epoch folding (EF, \citealt{leahy_searches_1983,leahy_searches_1987}) and $Z^2_n$ tests \citep{buccheri_search_1983} method implemented in the \texttt{Stingray} package. For \emph{XMM-Newton} observations of sources, we used photon events in the energy range 0.2-12.0\,keV from all instruments that were active and included the source of interest. These were screened to consider only those events that arrived at the detector when all instruments were considered to be in GTIs. For \emph{Chandra} observations of sources, we also considered photon events registered in the energy range 0.2-12.0\,keV by the ACIS instrument that were received at the detector during GTIs, when they were screened against the background. 

We performed blind searches for the periodicities in the photon event lists, in the frequency range from $4/T \leq f \leq 10^{-3}$\,Hz. As before, we selected this minimum frequency to ensure that there are at least four cycles contained within any single observation. The upper limit is chosen to reduce the probability of detecting false periodicities associated with the frame times of the instruments or with Poisson noise. Again, the frequency binning applied was $\Delta f = 1/25T$ and the exposures in phase bins were corrected to account for short breaks between GTIs. Candidate frequencies for the periodicities in these searches are identified as being significant at a $3\sigma$ level by means of a single-trial from multi-trial probability using the \texttt{fold\_detection\_level} and \texttt{z2\_n\_detection\_level} functions in \texttt{stingray}. 

We further assessed the significance of observed periodicities using simulated photon event lists. For each observation showing a significant periodicity candidate at the $3\sigma$ level as described above, in either the EF or $Z^2_n$ search, we simulated 2000 light curves. These light curves are simulated based on a PSD described by a power law, with a mean countrate and variance as based on the observation of interest. The index used for the power law is either that determined by fitting the PSD of the observation as described in Section \ref{subsec:meth_fourier}, or 1 if the PSD could not be well fitted with a power law. This index of 1 was selected, in the absence of a well fitted PSD, as being the most appropriate value following previous observations of $\gamma$ Cas analogues \citep[e.g.][]{robinson_search_2000,lopes_de_oliveira__2010}, which was further confirmed during this analysis (see below). We then determined the significance of the observed periodicity by comparing the observed EF or $Z^2_n$-statistic to the distribution of maximum values found in these 2000 simulations. The uncertainties on frequencies for observed periodicities are determined with one of two methodologies depending on how the peaks are presented in the EF or $Z^2_n$ statistic. If the peak is limited to a single, isolated, frequency bin then we report the uncertainty as the width of the frequency binning used in the periodicity search. If the peak is broad, encompassing several frequencies, we fit the peak in the EF or $Z^2_n$ statistic by a Gaussian curve. The uncertainty is reported as the FWHM of this peak. 
\vspace{-0.2cm}

\section{Frequency-dependent aperiodic variability}
\label{sec:freq_dep_var}

\begin{figure}
\centering
\includegraphics[width=0.95\columnwidth]{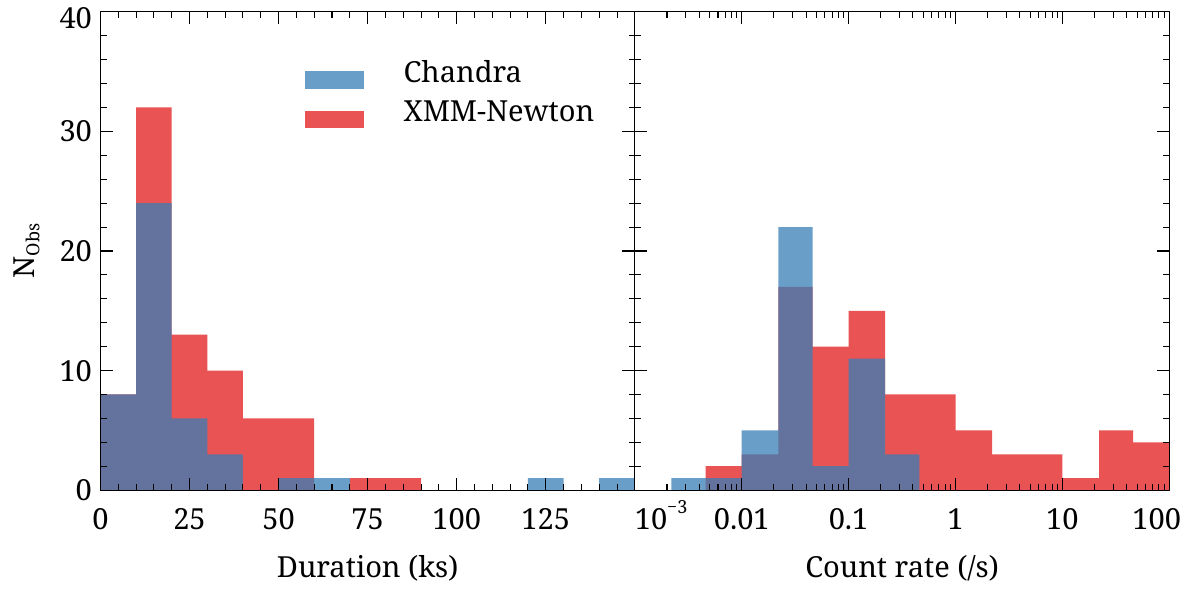}
  \caption{Distribution of observation lengths, and observed source average count rates in the \emph{Chandra} (blue) and \emph{XMM-Newton} EPIC (red) observations of $\gamma$ Cas and its analogues.}
     \label{fig:obs_exp_rate}
\vspace{-0.4cm}
\end{figure}

There were a total of 131 detections of the 26 identified $\gamma$ Cas analogues and 2 candidates, as listed in Table \ref{tab:all-obs}. In Figure \ref{fig:obs_exp_rate} we show the distribution of durations and average observed count rates across observations with the two observatories. As expected, due the different effective areas of the two observatories concerned, the \emph{XMM-Newton} observations recorded significantly higher average count rates than were observed in \emph{Chandra} observations. 

\begin{figure}
\centering
\includegraphics[width=0.75\columnwidth]{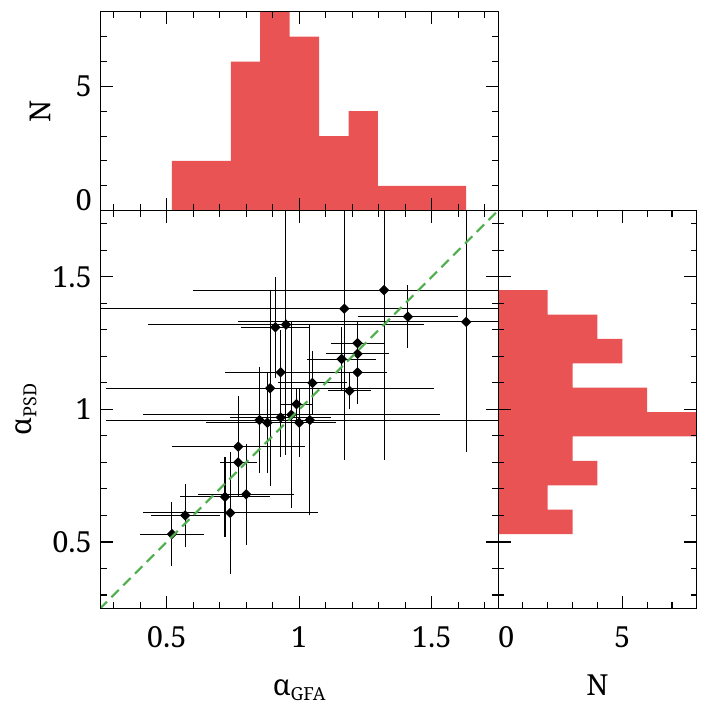}
  \caption{Indices derived from fitting a power law in PSDs of individual GTIs or GFAs of whole observations. The green dashed line represents a 1:1 relationship between the index from power-law fits to the GFA and PSD of a given observation. }
     \label{fig:FD_results}
\vspace{-0.5cm}
\end{figure}

It was possible to fit power-law relationships to the PSDs of light curves within GTIs in 36 observations, and to the GFAs of light curves in 35 exposures. This corresponds to at least one result for 18 of the 28 sources examined. The fitting to the PSDs and GFAs was only possible with a high degree of confidence for observations with \emph{XMM-Newton}, as the available \emph{Chandra} observations were too affected by Poisson noise to find broad enough sections of the PSD enabling a sensible fitting. There were four observations where multiple GTIs were longer than 10\,ks, each leading to a separate power-law fitting. The derived power-law indices ($PSD$ or $A^2 \propto f^{-\alpha}$) ranged $\alpha \sim0.5 - 1.5$ (see Table \ref{tab:res-aperiodic} and Fig. \ref{fig:FD_results}). This is in line with previous analyses of $\gamma$\,Cas analogues \citep[e.g.][]{frontera_time_1987,lopes_de_oliveira_x-ray_2007,lopes_de_oliveira__2010,naze_x-raying_2024}.

\begin{table*}
\centering
\caption{Aperiodic variability results, for $\gamma$ Cas analogues on top and candidates at the bottom.}
\label{tab:res-aperiodic}
\begin{tabular}{lcccc}
\hline\hline
Name & ObsID & $f$-Range (mHz) & $\alpha_{\text{PSD}}$ & $\alpha_{\text{GFA}}$ \\
\hline
\multirow{2}{*}{$\gamma$ Cas} & 0201220101 (1) & \multirow{2}{*}{$0.06-10.00$} & $0.95\pm0.07$ & \multirow{2}{*}{$0.91\pm0.06$} \\
 & 0201220101 (2) &  & $1.20\pm0.12$ &  \\
$\gamma$ Cas & 0651670201 & $0.23-10.00$ & $1.14\pm0.12$ & $1.22\pm0.11$\\
$\gamma$ Cas & 0651670301 & $0.26-10.00$ & $1.31\pm0.19$ & $0.91\pm0.13$ \\
$\gamma$ Cas & 0651670401 & $0.23-10.00$ & $1.21\pm0.11$ & $1.22\pm0.12$ \\
$\gamma$ Cas & 0651670501 & $0.18-10.00$ & $1.25\pm0.08$ & $1.22\pm0.10$ \\
$\gamma$ Cas & 0743600101 & $0.12-10.00$ & $1.07\pm0.07$ & $1.19\pm0.08$ \\
$\gamma$ Cas & 0840310101 & $0.24-10.00$ & $1.19\pm0.12$ & $1.16\pm0.13$ \\
$\gamma$ Cas & 0840310201 & $0.41-10.00$ & $1.14\pm0.16$ & $0.93\pm0.21$ \\
$\gamma$ Cas & 0840310301 & $0.64-10.00$ & $0.95\pm0.19$ & $0.88\pm0.23$ \\
$\gamma$ Cas & 0840310401 & $0.44-10.00$ & $1.35\pm0.12$ & $1.41\pm0.19$ \\
V782 Cas & 0102580801 & $0.13-1.00$ & -- & $1.06\pm0.43$ \\
$\zeta$ Tau & 0920020301 & $0.26-10.00$ & $1.10\pm0.12$ & $1.05\pm0.13$ \\
$\zeta$ Tau & 0920020401 & $0.20-5.00$ & $0.68\pm0.19$ & $0.80\pm0.18$ \\
HD44458 & 0820310301 & $0.32-2.00$ & $1.20\pm0.39$ & -- \\
HD45314 & 0670080301 & $0.16-1.50$ & -- & $1.00\pm0.31$ \\
HD90563 & 0880000601 & $0.19-1.50$ & -- & $1.09\pm0.29$ \\
\multirow{2}{*}{HD110432} & 0109480101 (1) & \multirow{2}{*}{$0.08-10.00$} & $0.73\pm0.12$ & \multirow{2}{*}{$0.85\pm0.06$} \\
 & 0109480101 (2) &  & $0.89\pm0.07$ & \\
HD110432 & 0109480201 & $0.09-10.00$ & $0.80\pm0.07$ & $0.77\pm0.07$ \\
\multirow{2}{*}{HD110432} & 0109480401 (1) & \multirow{2}{*}{$0.09-10.00$} & $0.82\pm0.10$ & \multirow{2}{*}{$0.78\pm0.07$} \\
 & 0109480401 (2) &  & $0.75\pm0.09$ & \\
\multirow{2}{*}{HD110432} & 0504730101 (1) & \multirow{2}{*}{$0.05-10.00$} & $1.05\pm0.08$ & \multirow{2}{*}{$0.90\pm0.05$} \\
 & 0504730101 (2) &  & $1.01\pm0.14$ & \\
HD110432 & 0840760201 & $0.08-10.00$ & $0.95\pm0.13$ & $1.00\pm0.14$ \\
HD119682 & 0087940201 & $0.11-2.00$ & $0.67\pm0.15$ & $0.72\pm0.17$ \\
HD119682 & 0551000201 & $0.76-3.00$ & $0.96\pm0.20$ & $0.85\pm0.16$ \\
HD119682 & 0840310901 & $0.37-1.00$ & $1.38\pm0.57$ & $1.17\pm1.27$ \\
HD119682 & 0840311001 & $0.18-1.00$ & $0.96\pm0.36$ & $1.04\pm0.77$ \\
HD119682 & 0840310801 & $0.38-2.00$ & $1.45\pm0.64$ & $1.32\pm0.72$ \\
V767 Cen & 0402121801 & $0.56-3.00$ & $1.08\pm0.30$ & -- \\
CQ Cir & 0840910401 & $1.08-5.00$ & $1.33\pm0.49$ & $1.63\pm0.86$ \\
HD157832 & 0551020101 & $0.20-2.50$ & $0.61\pm0.23$ & $0.74\pm0.33$ \\
HD157832 & 0810210301 & $0.14-2.50$ & $0.98\pm0.35$ & $0.97\pm0.56$ \\
HD161103 & 0691760101 & $0.19-3.00$ & $0.86\pm0.19$ & $0.77\pm0.25$ \\
V771 Sgr & 0840910501 & $0.26-10.00$ & $0.53\pm0.12$ & $0.52\pm0.12$ \\
V771 Sgr & 0886090801 & $0.20-10.00$ & $0.60\pm0.12$ & $0.57\pm0.13$ \\
SS397 & 0122700201 & $0.14-2.00$ & $0.73\pm0.23$ & -- \\
V558 Lyr & 0840200401 & $0.69-5.00$ & $1.08\pm0.37$ & $0.89\pm0.62$ \\
V2156 Cyg & 0840910601 & $0.28-2.00$ & $0.93\pm0.25$ & -- \\
$\pi$ Aqr & 0720390701 & $0.08-10.00$ & $1.02\pm0.06$ & $0.99\pm0.06$ \\
$\pi$ Aqr & 0932390901 & $0.49-10.00$ & $0.97\pm0.15$ & $0.93\pm0.19$ \\
\hline
HD42054 & 0402121401 & $0.29-2.00$ & $1.32\pm0.49$ & $0.95\pm0.52$ \\
\hline
\end{tabular}
\vspace{-0.25cm}
\tablefoot{Errors on the values for $\alpha$ are 1-$\sigma$ uncertainties.}
\vspace{-0.75cm}
\end{table*}

In accreting WDs, aperiodic variability is expected to be related to density fluctuations in the accretion disc onto non-magnetic WDs and somewhat magnetic (intermediate polar, IP) WDs \citep{balman_accretion_2020}. For accreting WDs with higher magnetic fields (polars), the magnetic field prevents the creation of an accretion disc, and so the aperiodic variability arises from density fluctuations in the polar accretion streams. In all CVs, the WD forms a close binary with a donor filling its Roche lobe. In symbiotic stars \citep{mukai_x-ray_2017,munari_symbiotic_2019}, the WDs are fed by stellar winds from red giant companions in larger orbits, with the variability resulting from the variable mass accretion rate from the winds. In CVs and symbiotic stars with lower magnetic fields, the aperiodic variability attributed to density fluctuations in the accretion disc has been well characterised by a broken power law of $PSD\propto f^{-1}$ at low frequencies with a break at $f\sim10$\,mHz and then $PSD\propto f^{-2}$ \citep{revnivtsev_observational_2011,balman_accretion_2020}.

The power-law exponents we recovered from fits to the PSDs of 40 GTIs range from --0.5 to --1.5 and are for frequencies below 10\,mHz. For 22 of the 40 PSDs, the exponent for the power-law fit is even fully consistent with --1 within the 1-$\sigma$ errors from fitting. For the remaining 18 cases a visual inspection indicates that an exponent of --1 cannot be excluded, as the derived PSDs are noisy (see Fig. \ref{fig:psd_fit_examples} for examples). The exponents derived from fits to the GFAs of 35 exposures range from --0.5 to --1.6, and are also for frequencies below 10\,mHz. We again find that for 20 of the 35 GFAs, the exponent for the power-law fit is fully consistent with --1 within the 1-$\sigma$ errors from fitting. As for the PSD fits, in the remaining 15 cases a visual inspection indicates that an exponent of --1 cannot be excluded. Although we do not consider the possibility of a broken power law in this work, this has been identified in a previous analysis of $\gamma$ Cas \citep{lopes_de_oliveira__2010}. The similarity in the behaviour of aperiodic variability is consistent with an accreting WD scenario for the $\gamma$\,Cas analogues \citep[as noted e.g. in][]{naze_x-raying_2024}. It would indicate the presence of an accretion disc, implying accreting WDs with moderate or no magnetic fields (the equivalent of non-magnetic CVs or IPs). This is in line with the latest X-ray results \citep{naze_xrism}.

\section{Periodic X-ray variability}
\label{sec:period}

\subsection{Detected periodicities}
\label{subsec:det_periods}

From the 40 PSD fits to power laws, GTIs in nine observations (of 6 different sources) contained periodicities, which were significant at the 3$\sigma$ level, using the test as described in Section \ref{subsec:meth_fourier}. The frequencies of these periodicities ranged from $f_{\text{PSD}}\sim0.27$ to 9.72\,mHz and there are no apparent trends between the power-law indices and the frequency values. This absence of trend is expected if periodicities are unrelated to the 'flickering' variability observed at higher frequencies. From the 131 observations of the $\gamma$ Cas analogues and candidates, there were significant peaks identified in the GFAs of 26 exposures of nine sources. They covered the frequency range from $f_{\text{GFA}}\sim0.09$ to 1.41\,mHz. In the nine observations where a periodic signal was detected in a single GTI, there was a detection of a periodic signal in the GFA analysis of the whole observation in seven cases. This was only at the same frequency in one case, observation 0840910501 of V771 Sgr, which was also the only occasion when a significant frequency was detected in PSD analysis below 1\,mHz. This difference in the detected periodicities is likely due to the implementation of oversampling applied to the GFA algorithm, which enables a more detailed exploration of the frequency range. These significant periodicities identified from the PSD and GFA analyses are listed in Table \ref{tab:res-periods}.

From the 131 observations we identified significant periodicities in 58 cases with the EF search and in 47 of these with the $Z^2_n$ search as well. These detections were all significant at a $3\sigma$ level on the basis of their EF or $Z^2_n$ statistic when accounting for the numbers of frequencies trialled. The detected frequencies covered the ranges $f_{\text{EF}}\sim0.06-0.98$\,mHz from the EF search, and $f_{Z^2_n}\sim0.09-0.98$\,mHz from the $Z^2_n$ search. The most significant periodicities identified by the blind searches with the EF and $Z^2_n$ statistic are listed in Table \ref{tab:res-periods}. 

\begin{table*}
\centering
\caption{Most significant periodicities detected in observations at least at a 3$\sigma$ significance level from PSDs, GFAs, or folded light curves (EF and $Z^2_n$ methods).}
\label{tab:res-periods}
\begin{tabular}{lcccccccr}
\hline
Name  &  ObsID  &  Exp.  &  $f_{\text{PSD}}$  &  $f_{\text{GFA}}$  &  $f_{\text{EF}}$  &  $f_{Z^2_n}$  \\
  &    &  (ks)  &  (mHz)  &  (mHz)  &  (mHz)  &  (mHz)  \\
\hline
$\gamma$ Cas  & 0201220101 & 71 &  --  &  $\mathbf{0.10\pm0.01}$  &  $0.07\pm0.02$  &  $\mathbf{0.10\pm0.01}$  \\
$\gamma$ Cas  & 0651670201 & 17.9 &  $5.01\pm0.04$  &  $\mathbf{0.58\pm0.05}$  & $0.29\pm0.04$ &  $\mathbf{0.59\pm0.05}$ \\
$\gamma$ Cas  & 0651670301 & 16.1 &  $4.71\pm0.08$  &  $\mathbf{0.32\pm0.08}$ &  $\mathbf{0.28\pm0.07}$  &  $0.85\pm0.07$  \\
$\gamma$ Cas  & 0651670401 & 17.9 &  --  &  $\mathbf{0.45\pm0.05}$  &  $0.23\pm0.15$  & $\mathbf{0.45\pm0.05}$  \\
$\gamma$ Cas  & 0651670501 & 23.8 &  --  &  $0.29\pm0.04$  &  $0.24\pm0.03$  &  $0.34\pm0.04$  \\
$\gamma$ Cas  & 0743600101 & 34 &  $7.02\pm0.02$  &  $\mathbf{0.15\pm0.03}$  &  $\mathbf{0.14\pm0.02}$  &  $\mathbf{0.15\pm0.03}$  \\
$\gamma$ Cas  & 0840310101 & 19.9 &  --  &  $\mathbf{0.25\pm0.04}$  &  $\mathbf{0.26\pm0.06}$ &  $\mathbf{0.25\pm0.04}$  \\
$\gamma$ Cas  & 0840310201 & 18.5 &  --  &  $\mathbf{0.44\pm0.09}$  &  $\mathbf{0.44\pm0.10}$  &  $\mathbf{0.43\pm0.08}$  \\
$\gamma$ Cas  & 0840310301 & 9.9 &  --  &  $\mathbf{0.98\pm0.18}$  &  $0.67\pm0.09$  &  $\mathbf{0.98\pm0.22}$  \\
$\gamma$ Cas  & 0840310401 & 17.6 &  --  &  $\mathbf{0.44\pm0.05}$  &  $\mathbf{0.45\pm0.04}$  &  $\mathbf{0.45\pm0.04}$  \\
TYC 3681-695-1  & 0112200201 & 9 &  --  &  --  & $\mathbf{0.98\pm0.12}$  &  $\mathbf{0.96\pm0.14}$  \\
V782 Cas  & 0102580801 & 50.4 &  --  &  --  &  $\mathbf{0.22\pm0.04}$  &  $\mathbf{0.22\pm0.02}$  \\
$\zeta$ Tau  & 0920020301 & 17 &  --  &  $\mathbf{0.27\pm0.07}$  & $\mathbf{0.27\pm0.16}$ &  $\mathbf{0.27\pm0.06}$  \\
$\zeta$ Tau  & 0920020401 & 22 &  --  &  $\mathbf{0.41\pm0.04}$ &  $0.21\pm0.04$  &  $\mathbf{0.41\pm0.03}$  \\
HD44458  & 0820310301 & 15.9 &  --  &  $\mathbf{0.37\pm0.07}$  &  $\mathbf{0.35\pm0.11}$  &  $\mathbf{0.37\pm0.07}$  \\
HD45314  & 0670080301 & 26.2 &  --  &  --  &  $0.18\pm0.01$  &  $0.31\pm0.03$  \\
HD45314  & 0760220601 & 31 &  --  &  --  &  $\mathbf{0.19\pm0.03}$  &  $\mathbf{0.18\pm0.03}$  \\
HD45995  & 0820310401 & 13 &  --  &  --  &  $\mathbf{0.82\pm0.36}$  &  $\mathbf{0.82\pm0.11}$  \\
HD90563  & 0860650501 & 49 &  --  &  --  &  $\mathbf{0.11\pm0.03}$  &  $\mathbf{0.11\pm0.02}$  \\
HD110432  & 0109480101 & 53 &  $2.55\pm0.04$  &  $\mathbf{0.09\pm0.02}$  & $\mathbf{0.09\pm0.02}$  &  $\mathbf{0.09\pm0.02}$  \\
HD110432  & 0109480201 & 48.7 &  --  &  $\mathbf{0.14\pm0.02}$  &  $\mathbf{0.14\pm0.02}$ &  $\mathbf{0.14\pm0.02}$  \\
HD110432  & 0109480401 & 48 &  --  & $\mathbf{0.12\pm0.02}$  & $\mathbf{0.12\pm0.03}$ &  $\mathbf{0.12\pm0.02}$  \\
HD110432  & 0504730101 & 80.7 &  --  & $\mathbf{0.11\pm0.01}$ & $\mathbf{0.06\pm0.07}$  & $\mathbf{0.11\pm0.01}$ \\
HD110432  & 0840760201 & 17 &  $9.72\pm0.04$  & $\mathbf{0.26\pm0.10}$ &  $\mathbf{0.27\pm0.06}$ & $\mathbf{0.27\pm0.05}$ \\
HD119682  & 0087940201 & 40.5 &  --  & $\mathbf{0.22\pm0.03}$ & $\mathbf{0.11\pm0.31}$  &  $\mathbf{0.23\pm0.04}$ \\
HD119682  & 0551000201 & 57.3 &  --  &  $0.31\pm0.01$  & $\mathbf{0.10\pm0.02}$ &  $\mathbf{0.10\pm0.02}$ \\
HD119682  & 0840310901 & 14.3 &  --  &  --  &  $0.39\pm0.04$  &  $0.79\pm0.09$  \\
HD119682  & 0840311001 & 25.3 &  --  &  --  & $\mathbf{0.19\pm0.02}$ &  $\mathbf{0.19\pm0.02}$ \\
HD119682  & 0840311101 & 14.5 &  --  &  --  & $\mathbf{0.39\pm0.09}$ & $\mathbf{0.39\pm0.08}$ \\
HD119682  & 0840310801 & 14.2 &  $1.68\pm0.08$  &  --  &  $\mathbf{0.42\pm0.07}$  & $\mathbf{0.43\pm0.08}$ \\
V767 Cen  & 0402121801 & 11 &  --  &  --  &  $0.57\pm0.07$  &  --  \\
V767 Cen  & 0891800801 & 12 &  --  &  --  &  $0.79\pm0.21$  &  --  \\
HD157832  & 0551020101 & 21.9 &  --  &  --  &  $0.20\pm0.03$  &  $0.46\pm0.04$  \\
HD157832  & 0810210301 & 30.4 &  --  &  --  & $\mathbf{0.14\pm0.04}$ &  $\mathbf{0.14\pm0.03}$  \\
HD161103  & 0201200101 & 17.8 &  --  &  --  & $\mathbf{0.26\pm0.05}$ &  $\mathbf{0.28\pm0.04}$ \\
HD161103  & 0691760101 & 22.9 &  --  &  --  & $\mathbf{0.21\pm0.04}$ & $\mathbf{0.21\pm0.05}$ \\
V771 Sgr  & 0840910501 & 19.8 & $\mathbf{0.27\pm0.04}$ & $\mathbf{0.26\pm0.07}$   & $\mathbf{0.26\pm0.04}$ & $\mathbf{0.26\pm0.03}$ \\
V771 Sgr  & 0886090801 & 23 &  --  &  $\mathbf{0.49\pm0.04}$  &  $0.20\pm0.05$  &  $\mathbf{0.49\pm0.04}$  \\
HD316568  & 0402280101 & 44.1 &  --  &  --  &  $0.57\pm0.03$  &  --  \\
GSC2 S300302371  & 0135741001 & 8.5 &  --  &  --  &  $0.85\pm0.03$  &  --  \\
SS397  & 1233 & 14 &  --  &  $1.41\pm0.06$  &  $0.35\pm0.02$  &  --  \\
SS397  & 0122700301 & 34.2 &  --  &  --  &  $0.15\pm0.01$  &  $0.62\pm0.03$  \\
SS397  & 0122700401 & 32.9 &  --  &  --  &  $0.23\pm0.02$  &  --  \\
SS397  & 0122700501 & 32.6 &  --  &  --  &  $0.30\pm0.02$  &  --  \\
SS397  & 0122700801 & 17.4 &  --  &  --  &  $0.42\pm0.04$  &  $0.50\pm0.05$  \\
SS397  & 0890200301 & 13 &  --  &  --  &  $0.42\pm0.04$  &  --  \\
NGC 6649 9  & 0122700101 & 35.3 &  --  &  --  &  $\mathbf{0.14\pm0.04}$ &  $\mathbf{0.14\pm0.03}$  \\
NGC 6649 9  & 0122700401 & 32.9 &  --  &  --  &  $\mathbf{0.16\pm0.02}$ &  $\mathbf{0.16\pm0.02}$  \\
NGC 6649 9  & 0890200601 & 13 &  --  &  --  &  $0.52\pm0.04$  &  --  \\
NGC 6649 9  & 0890200901 & 21.8 &  --  &  --  &  $0.36\pm0.01$  &  --  \\
3XMM J1901+0459  & 0840910801 & 39.9 &  --  &  --  &  $\mathbf{0.13\pm0.08}$ &  $\mathbf{0.14\pm0.05}$  \\
V558 Lyr  & 0840200401 & 9 &  $3.30\pm0.09$  &  --  &  $0.75\pm0.08$  &  --  \\
SAO 49725  & 0201200201 & 25.9 &  --  &  --  & $\mathbf{0.63\pm0.10}$ & $\mathbf{0.63\pm0.07}$ \\
V2156 Cyg  & 0840910601 & 18 &  --  &  --  & $\mathbf{0.28\pm0.04}$ & $\mathbf{0.28\pm0.04}$ \\
$\pi$ Aqr  & 0720390701 & 54.4 &  --  & $\mathbf{0.10\pm0.02}$ &  $0.08\pm0.01$  & $\mathbf{0.10\pm0.01}$ \\
$\pi$ Aqr  & 0932390901 & 10 &  --  & $\mathbf{0.63\pm0.10}$  & $\mathbf{0.63\pm0.14}$ & $\mathbf{0.63\pm0.09}$ \\
\hline
HD42054  & 0402121401 & 15.7 &  $1.40\pm0.06$  & $\mathbf{0.32\pm0.08}$ &  $\mathbf{0.32\pm0.06}$ & $\mathbf{0.32\pm0.06}$ \\
HD120678  & 0820310601 & 50 &  --  &  --  & $\mathbf{0.10\pm0.02}$  &  $\mathbf{0.10\pm0.01}$ \\
\hline
\end{tabular}
\tablefoot{The columns $f_{\text{PSD}}$, $f_{\text{GFA}}$, $f_{\text{EF}}$, and $f_{Z^2_n}$ give the periodic frequency. Detections within observationsthat are consistent within errors across methods are highlighted in \textbf{bold}.}
\end{table*}

From our PSD, GFA, EF, and $Z^2_n$ searches we recovered (within 1-$\sigma$ errors) the previously reported periodicities in \emph{XMM-Newton} and \emph{Chandra} observations of $\gamma$ Cas \citep{lopes_de_oliveira__2010}, $\zeta$ Tau, V771 Sgr, and HD161103. We did not recover the periodicity identified by \citet{huenemoerder_chandra_2024} in $\pi$ Aqr although we detected a harmonic of this signal (see Sect. \ref{subsec:discuss_piaqr} below).

The disagreement in periodicities detected across the GFA, EF, and $Z^2_n$ approaches in some observations could be the result of spurious detections or of the different assumptions of each approach. The GFA and $Z^2_n$ approaches both assume a sinusoidal shape to any periodicity, with the possibility of harmonics included in the profile expected by the $Z^2_n$ search, while the EF approach makes no assumptions with respect to the profile shape of any periodic variability. The fact that there is no assumption on the shape of any periodicity for the EF search might also explain why EF detections may not be found through PSD or GFA searches. Additionally, our implementation of the GFA uses the light curves with photon data being binned previously, whereas the EF and $Z^2_n$ approaches both use the raw event arrival times. In several cases, this resulted in the identification of harmonics of a common frequency with the different approaches in some observations. As an example, the detected periodicity frequency in observation 0651670201 of $\gamma$ Cas identified by the EF search is exactly half of that identified by both the GFA and $Z^2_n$ methods and a visual inspection of the folded light curves confirms that method difference is indeed the cause. Furthermore, it is also notable that in Table \ref{tab:res-periods} we only report the most significant feature detected by the GFA, EF and $Z^2_n$ methods. For example, in observation 0087940201 of HD119682 the signal detected by the $Z^2_n$ and GFA searches at 0.23\,mHz also appears in the EF search with a greater than $3\sigma$ significance, but it has a lower significance than the main signal at 0.11\,mHz. Varying frequencies across approaches might therefore comprise an artifact and a more detailed individual analysis is thus needed.

As a final remark, we note that there were no \emph{Chandra} observations resulting in any significant detections of periodicities through the PSD, EF, or $Z^2_n$ searches. There was only one observation that identified a significant periodicity through the GFA analyses. The exposures of many \emph{Chandra} observations were of sufficient length to permit the exploration of a broad frequency range, but the observed count rates are lower than in comparable \emph{XMM-Newton} observations, as shown in Figure \ref{fig:obs_exp_rate}. Thus, it appears unlikely for us to be able to detect any periodicities in \emph{Chandra} observations as a result of this lower sensitivity and the subsequent reduction in the S/N of any periodic variability. Therefore, we have chosen to limit our discussion to the persistence of periodicity in the following section with respect to the available \emph{XMM-Newton} observations of the $\gamma$ Cas analogues and candidates.

\subsection{Examining whether the detected periodicities are persistent}
\label{subsec:period_best}

\begin{figure}[h]
\centering
\includegraphics[width=0.85\linewidth]{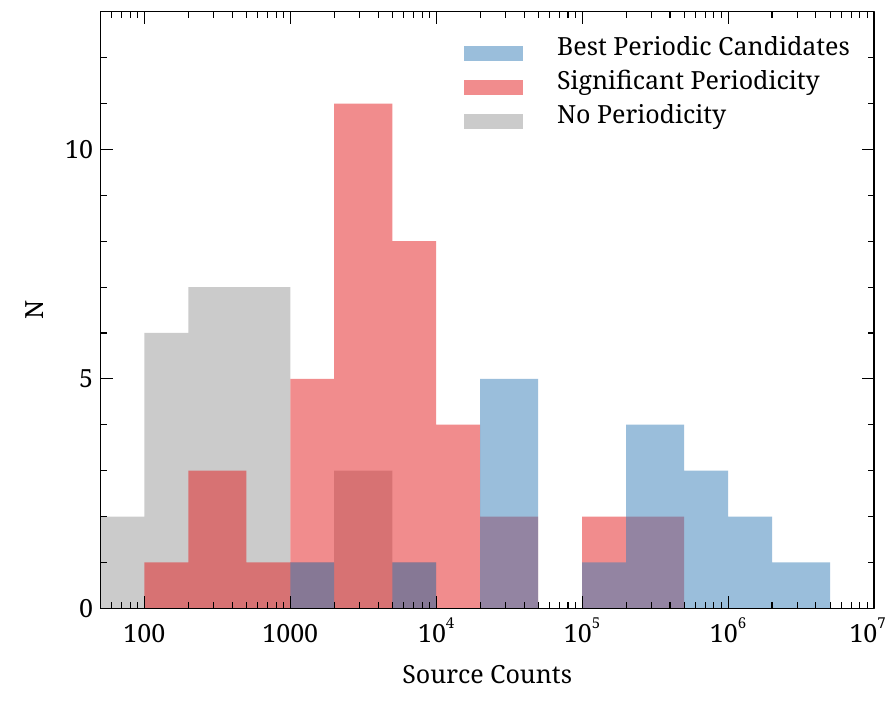}
  \caption{Histograms showing populations of source photon counts in \emph{XMM-Newton} observations with and without detections of significant periodicities. The observations classified as the 'best periodic candidates'  characterise the five candidates exhibiting significant and persistent periodicities across observations: $\gamma$ Cas, $\zeta$ Tau, HD45314, V771 Sgr, and $\pi$ Aqr.}
     \label{fig:rates_vs_dets}
\vspace{-0.25cm}
\end{figure}

Observations vary in length and the targets display various count rates: all of these parameters can affect the detectability of a signal. For example, significant periodicities clearly appear more often detected in sources with a higher number of source counts (Fig. \ref{fig:rates_vs_dets}). This highlights the need for high-quality pointed observations to provide the data necessary for signal detection. Despite these difficulties linked to the dataset variety, we made a concerted effort to assess the reality of the detected signals.

From the 28 sources examined, there were only three where significant periodicities were not detected in any of their observations: CQ Cir, 2XMM J180816.6-191939, and V810 Cas (which had no \emph{XMM-Newton} observations and only one \emph{Chandra} observation). While 2XMM J180816.6-191939 is a rather faint source, this is not the case of the other two, but the exposures were not very long (7-12\,ks), which could stand in the way of detecting periodicities, as reported in other cases deemed as 'persistent'.

For 8 of the remaining 25 sources, there was only one available \emph{XMM-Newton} observation, and a significant periodicity was detected in each of these observations: TYC 3681-695-1, V782 Cas, HD45995, HD45995, V558 Lyr, SAO 49725, V2156 Cyg, and HD42054 ($\gamma$ Cas analogue candidate). The signals were identified at the same frequency in both the EF and $Z^2_n$ searches for TYC 3681-695-1, V782 Cas, HD45995, SAO49725, V2156 Cyg, and in the EF, $Z^2_n$  and GFA analyses of HD44458 and HD42054. A visual inspection of the profile of the periodicity identified by the EF search in the observation of the last target, V558 Lyr, showed that it was far from sinusoidal in shape, explaining the non-detections by the GFA and $Z^2_n$ approaches in that case. For these 8 sources, no conclusion as to the persistence of the periodicities can be drawn, as only a single observation is available. Additional exposures will be required to confirm their presence.

There were seven sources with multiple available \emph{XMM-Newton} observations and with a significant periodicity detected in at least one case (but not in all datasets), namely: HD90563, HD316568, GSC2 S300302371, SS397, NGC 6649 9, 3XMM J1901+0459, and HD120678 ($\gamma$ Cas analogue candidate). The periodicity of HD120678 was detected on the longest observation available, which is a priori the most sensitive dataset. The other observations are ten times shorter hence the lack of detection in these exposures might be an observational artifact due to a lower S/N. Thus, no conclusions can be drawn with respect to the potential persistence of this periodicity; thus, it remains a candidate periodicity and can only be confirmed via subsequent observations. In contrast, the periodicities of HD316568 and GSC2 S300302371 were detected in only one observation, but considering the observational characteristics (exposure length and source count rate), such signals should have been detected in the other observations of these sources if they do indeed correspond to a persistent variability. The lack of detections therefore suggests that they are potentially spurious. In addition, among the periodicities detected in different observations of SS397 and NGC 6649, nine of them have different frequencies and different patterns revealed by the folded light curves; in addition, several observations do not yield detections while they would otherwise be expected to. We therefore conclude that the identified periodicities in these two sources are most likely to be spurious. Finally, the periodicities identified in a single observation of HD90563 and 3XMM J1901+0459 appear beyond the frequency ranges initially probed for their other observations, as these observations are shorter; thus, there would be fewer than four cycles of the identified periodicities in these exposures. However, contrary to the case of HD120678, the exposure lengths are not widely different, so the noise would a priori be just slightly higher in them. As a check, we therefore decided to broaden the searches to lower frequencies. Such a search for periodicities in the range $f=0.1-0.2$\,mHz in the sole other observation of 3XMM J1901+0459 (ObsID 0136030201) revealed no significant periodicities; meanwhile a search in the range $f=0.100-0.125$ in the five other observations of HD90563 only identified one significant periodicity at $f_{\text{EF}}=0.11$ in one observation (ObsID 0880000501). Thus, there is no evidence that the detected periodicities are persistent in either source; hence, their detections might have been spurious.

The remaining ten sources all showed significant periodic frequencies in all of their \emph{XMM-Newton} observations: $\gamma$ Cas, $\zeta$ Tau, HD45314, HD110432, HD119682, V767 Cen, HD157832, HD161103, V771 Sgr, and $\pi$ Aqr. In one case, HD45314, the frequency identified is persistent across observations, as the most significant periodic frequency identified by the EF and $Z^2_n$ searches. For the other nine sources, there is no immediately evident persistent periodicity across observations, as the most significant frequency identified by the GFA, EF, and $Z^2_n$ searches is not consistent. This may be due to the varying lengths of observations (hence different frequency ranges explored, with a detected signal sometimes outside them) or due to the identification of harmonics or other signals as more significant features. To clarify the situation, we conducted additional searches targeted on frequencies near the identified signals. Such targeted searches for HD110432, HD119682, V767 Cen, HD157832, and HD161103 revealed no consistent or persistent periodic frequencies across any of the observations. The varying signals found in these five sources will thus require further observation to confirm their nature. The other five sources, $\gamma$ Cas, $\zeta$ Tau, HD45314, V771 Sgr and $\pi$ Aqr, display persistent periodic variability, which we summarise in the following sections. Below, we report the significances of detections that take into account the ranges of frequencies tested. The robustness of detections of periodicities in all of the sources considered are summarised in Table \ref{tab:period-class}.

\begin{table}
\centering
\caption{Robustness of detected periodicities.}
\label{tab:period-class}
\begin{tabular}{lc}
\hline\hline
Class & Sources \\
\hline
\multirow{2}{*}{No detection} & CQ Cir, V810 Cas, \\
 & 2XMM J180816.6-191939 \\
 \hline
\multirow{3}{*}{Spurious} & HD90563, HD316568, SS397, \\
 & GSC2S300302371, NGC6649 9, \\
 & 3XMM J1901+0459, HD110432, \\
 & HD119682, V767 Cen, HD157832, HD161103 \\
\hline
\multirow{3}{*}{Candidate} & TYC 3681-691-1, V782 Cas, HD44458, \\
& HD45995, V558 Lyr, SAO 49725, \\
& V2156 Cyg, HD42054, HD120678\\
\hline
\multirow{2}{*}{Persistent} & $\gamma$ Cas, $\zeta$ Tau, HD45314, \\
 & V771 Sgr, $\pi$ Aqr \\
\hline
\end{tabular}
\tablefoot{Sources are classified as: no detection, potentially spurious detection, candidate periodicity (to be confirmed), and persistent detection.}
\vspace{-0.25cm}
\end{table}

\subsubsection{$\gamma$ Cas}
\label{subsec:discuss_gcas}
\begin{figure}[h]
\centering
\includegraphics[width=0.925\linewidth]{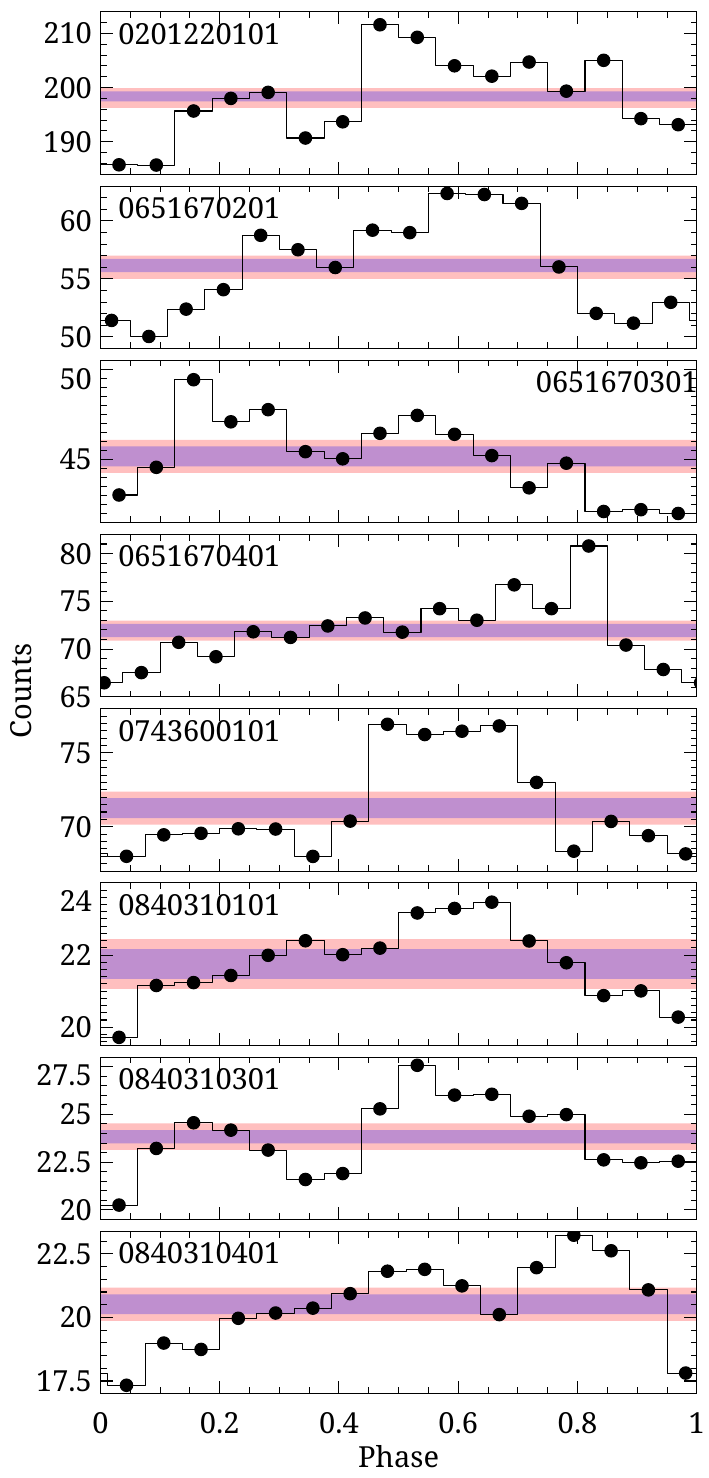}
  \caption{Light curve of $\gamma$ Cas folded using a period of $P\sim1.7$\,ks and binned using 16 bins. Folding periods are as per the values in Table \ref{tab:gam-cas-per-grps}. The purple and pink bands denote the $3\sigma$ and $5\sigma$ Poissonian confidence intervals around the mean value. The value of $T_0$ is arbitrarily selected per observation, so that the minimum of the periodicity profile is set at phase 0.}
     \label{fig:gcas_pulse1}
\vspace{-0.25cm}
\end{figure}

\begin{figure}[h]
\centering
\includegraphics[width=0.925\linewidth]{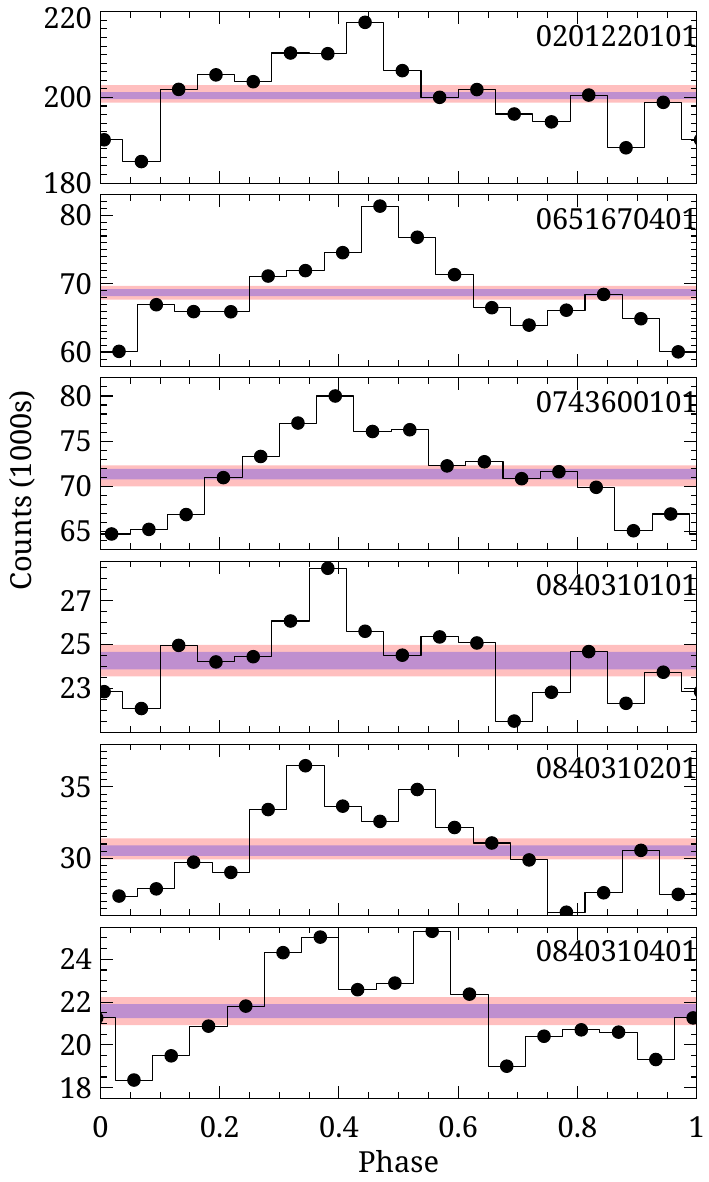}
  \caption{Same as Fig. \ref{fig:gcas_pulse1} but for a period of $P\sim2.5$\,ks.  }
     \label{fig:gcas_pulse2}
\vspace{-0.25cm}
\end{figure}

The frequencies detected in the ten \emph{XMM-Newton} observations vary significantly across observations and they do not always agree within the acceptable errors. However, partial agreement is observed. Indeed, three observations (0651670401, 0840310201, 0840310401) showed a consistent periodicity $f\sim0.45$\,mHz ($P\sim2.3$\,ks), while the periodicities detected in some of the other observations are harmonics of this frequency. It is possible that the signal is also present in other exposures, but it might be stronger in one, compared to another, or it could simply lie beyond the initial frequency range. To avoid these problems, we performed a targeted search in the range $f=0.33-0.67$\,mHz ($P = 1.5-3.0$\,ks). Two groups of periodicities were then identified at frequencies $f\sim 0.6$\,mHz and $f\sim 0.4$\,mHz (or $P\sim1.7$\,ks and $P\sim2.5$\,ks) all with a significance of $>5\sigma$ as per the $\chi^2$ test for periodicity against the EF and $Z^2_n$ statistic \citep{leahy_searches_1983,buccheri_search_1983,leahy_searches_1987,huppenkothen_stingray_2019}, as shown in Table \ref{tab:gam-cas-per-grps}. The recurrent detections indicate it is unlikely for the signals to be spurious and the overall significance of these detections, even considering the large number of observations (ten), is greater than $5\sigma$. However, we note that the period values are not exactly identical nor are the shapes of the folded and binned light curves (as shown in Fig. \ref{fig:gcas_pulse1} and Fig. \ref{fig:gcas_pulse2}).

\begin{figure}[h]
\centering
\includegraphics[width=0.9\linewidth]{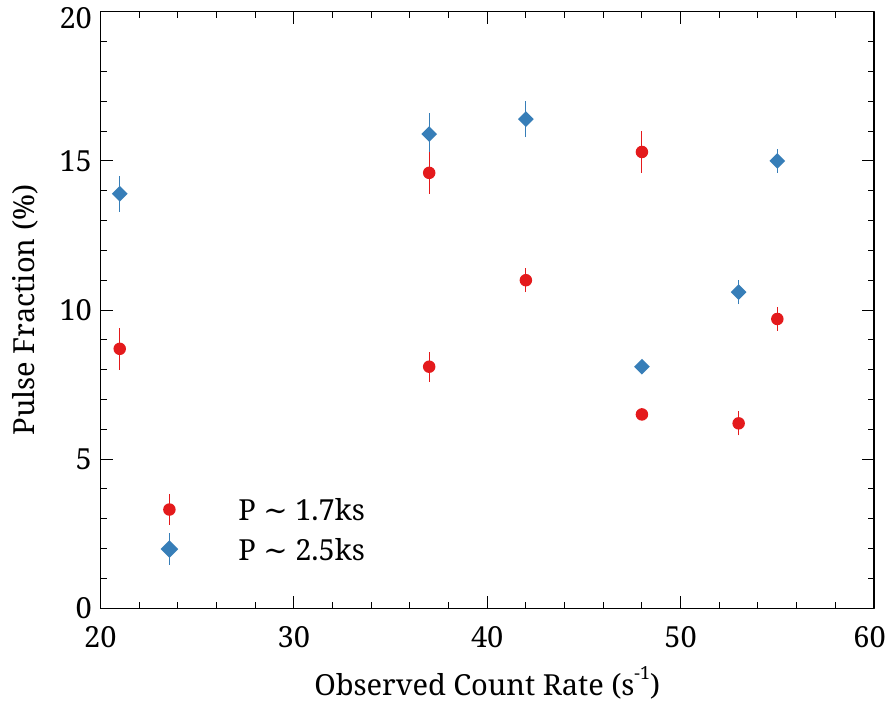}
  \caption{Pulsed fraction for periodicities of $P\sim1.7$\,ks (red circle) and $P\sim2.5$\,ks (blue diamond) detected across observations of $\gamma$ Cas compared with the observed source count rate.}
     \label{fig:gc_pf_rate}
\end{figure}

To characterise the signal strength, we used the pulsed fraction, defined as $PF = (A_{\text{max}} - A_{\text{min}})/(A_{\text{max}} + A_{\text{min}})$, where $A_{\text{max}}$ and $A_{\text{min}}$ are the maximum and minimum counts of the folded and binned light curve. It lies in the 6--15\% and 8--16\% range for the signals at $P\sim1.7$\,ks and $P\sim2.5$\,ks, respectively. X-ray observations were taken when the Be star showed various disc states, as traced by H$\alpha$ strengths and visual magnitudes \citep{rauw_x-ray_2022}. However, there is no apparent correlation between the detection of a periodicity or its strength and the X-ray brightness (see Fig \ref{fig:gc_pf_rate}) or the evolution of the Be star decretion disc \citep{rauw_x-ray_2022}. 

\begin{table*}
\centering
\caption{Significant (at $\geq3\sigma$ level) periodicities  detected in the range $f=0.33-0.67$\,mHz in observations of $\gamma$ Cas.}
\label{tab:gam-cas-per-grps}
\begin{tabular}{lcccccccr}
\hline\hline
ObsID & Exp & Rate & $f_1$ & $P_1$ & PF$_1$ & $f_2$ & $P_2$ & PF$_2$ \\
 & (ks) & (cts $s^{-1}$) & (mHz) & (ks) & (\%) & (mHz) & (ks) & (\%) \\
\hline
0201220101 & 71 & 48 & $0.56\pm0.02$ & $1.77\pm0.07$ & $6.5\pm0.2$ & $0.36\pm0.01$ & $2.77\pm0.09$ & $8.1\pm0.2$ \\
0651670201 & 18 & 42 & $0.59\pm0.05$ & $1.70\pm0.14$ & $11.0\pm0.4$ & -- & -- & -- \\
0651670301 & 16 & 37 & $0.59\pm0.09$ & $1.70\pm0.23$ & $8.1\pm0.5$ & -- & -- & -- \\
0651670401 & 18 & 55 & $0.58\pm0.04$ & $1.72\pm0.12$ & $9.7\pm0.4$ & $0.45\pm0.05$ & $2.20\pm0.25$ & $15.0\pm0.4$ \\
0651670501 & 24 & 37 & -- & -- & -- & -- & -- & -- \\
0743600101 & 34 & 53 & $0.56\pm0.03$ & $1.78\pm0.11$ & $6.2\pm0.4$ & $0.41\pm0.03$ & $2.41\pm0.16$ & $10.6\pm0.4$ \\
0840310101 & 20 & 21 & $0.66\pm0.07$ & $1.50\pm0.15$ & $8.7\pm0.7$ & $0.38\pm0.04$ & $2.63\pm0.27$ & $13.9\pm0.6$ \\
0840310201 & 19 & 42 & -- & -- & -- & $0.43\pm0.08$ & $2.33\pm0.45$ & $16.4\pm0.6$ \\
0840310301 & 10 & 48 & $0.51\pm0.09$ & $1.98\pm0.37$ & $15.3\pm0.7$ & -- & -- & -- \\
0840310401 & 18 & 3& $0.67\pm0.05$ & $1.50\pm0.12$ & $14.6\pm0.7$ & $0.42\pm0.07$ & $2.39\pm0.41$ & $15.9\pm0.7$ \\
\hline
\end{tabular}
\tablefoot{Columns from left to right: observation ID, total exposure, periodicity frequency $f_1$ (if detected) at approximately $f\sim0.6$\,mHz, the corresponding period ($P_1$), pulsed fraction of this periodicity PF$_1$, periodicity frequency $f_2$ (if detected) at approximately $f\sim0.4$\,mHz, the corresponding period ($P_2$), and the pulsed fraction of this periodicity, PF$_2$.}
\vspace{-0.25cm}
\end{table*}

The presence of two periodicities can appear puzzling if a periodic signal is assumed to be linked to WD rotation. However, a pair of signals could arise from a beating effect, caused by the interference of two signals with $f_{\text{beat}}=f_1 - f_2$. For example, if we take the periodicity $P\sim1.7$\,ks as the main rotation variability, then the periodicity at $P\sim2.3$\,ks (taken as a so-called beat periodicity) can be created from the interference of the rotation with a second signal at $P\sim6.5$\,ks. With the X-rays from $\gamma$ Cas being consistent with that of accreting magnetic WDs \citep{naze_xrism}, the longer period could be associated with material orbiting the WD in a truncated accretion disc as typically seen in IPs \citep[e.g.][]{revnivtsev_observational_2011,luna_constraining_2018}. For a WD companion with $M=1M_{\odot}$ and $R_{\text{WD}}=0.012R_{\odot}$, a periodicity of $P\sim6.5$\,ks associated with Keplerian rotation would correspond to a radius of $r\sim63R_{\text{WD}}$, inside the Roche lobe of the WD ($\sim68R_{\odot}$, \citealt{rauw_multi-faceted_2025}). Alternatively, these could be X-ray counterparts to the optical QPOs seen in a small number of accreting WDs that exhibit harmonics \citep{veresvarska_discovery_2024}.

\subsubsection{$\zeta$ Tau}
\label{subsec:discuss_ztau}

\begin{figure}[h]
\centering
\includegraphics[width=0.9\linewidth]{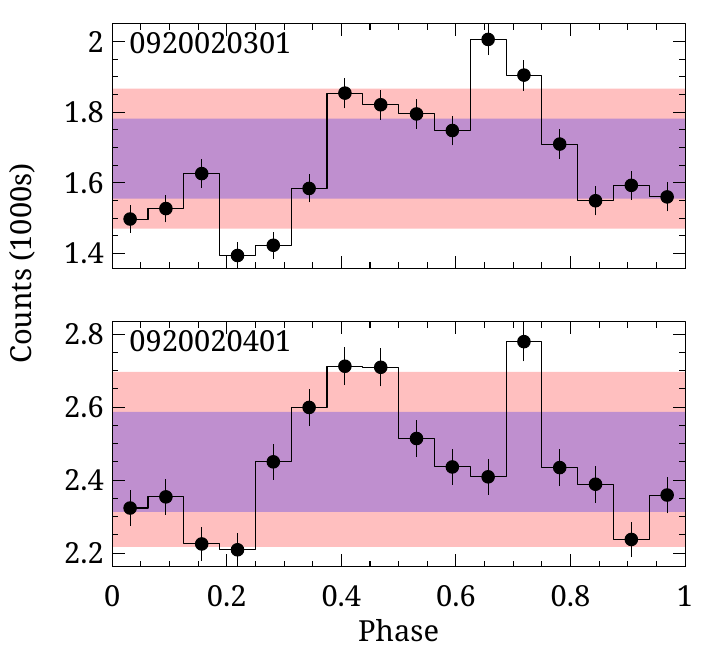}
  \caption{Same as Fig. \ref{fig:gcas_pulse1} but for $\zeta$ Tau. The periodicity used for folding observation 0920020301 is 1.28\,ks while it is 1.14\,ks for 0920020401.}
     \label{fig:ztau_pulse}
\vspace{-0.25cm}
\end{figure}

A prior analysis of the \emph{XMM-Newton} observations of $\zeta$ Tau \citep{naze_x-raying_2024} identified a potential periodicity of $P\sim3.7$\,ks in the first of the two observations used here, which we recovered with significance $>5\sigma$. In the second observation, we identified a periodicity at $P=2.4$\,ks in the GFA and $Z^2_n$ searches with a significance of $>5\sigma$. These periods are not consistent with one another, however, we note that they could both be harmonics of a shorter period, $P\sim1.2$\,ks. To assess this possibility, we performed a targeted search in the range $f=0.77-0.91$\,mHz, which identified signals at $P=1.28\pm0.07$\,ks ($f=0.78\pm0.04$\,mHz) and $P=1.14\pm0.10$\,ks ($f=0.88\pm0.07$\,mHz) in the first and second \emph{XMM-Newton} observations, both with a significance of $>5\sigma$ (Fig. \ref{fig:ztau_pulse}). The significance of these detections when considering the two observations is greater than $5\sigma$. Their pulsed fractions are 18\% and 12\% in the first and second observations, respectively. In other words, the pulsed fraction appears higher when the source is fainter. The pulsed fraction is slightly higher (although consistent within errors) in both observations in the soft band (0.2--3.0\,keV) of 20\% and 17\%, against a hard band (3.0--12.0\,keV), of 18\% and 12\%.
\vspace{-0.25cm}

\subsubsection{HD45314}
\label{subsec:discuss_hd45314}

\begin{figure}
\centering
\includegraphics[width=0.9\linewidth]{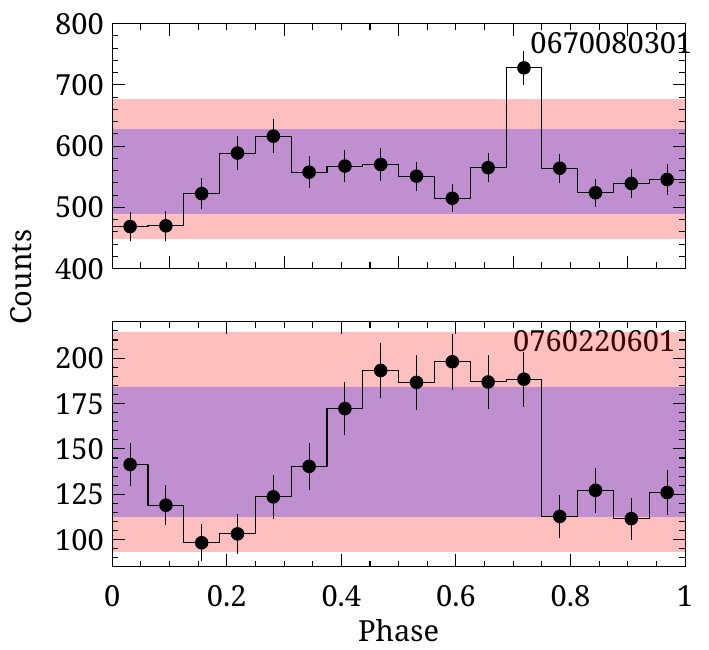}
  \caption{Same as Fig. \ref{fig:gcas_pulse1} but for HD45314. The periodicity used for folding observation 0670080301 is 5.48\,ks while it is 5.43\,ks in 0760220601.}
     \label{fig:HD43514_pulse}
\vspace{-0.25cm}
\end{figure}

In the two observations of HD45314 with \emph{XMM-Newton}, we retrieved a persistent periodicity of $P\sim5.4\pm0.9$\,ks ($f=0.18\pm0.03$\,mHz) with the epoch folding methods (i.e. EF for the first observation, EF and $Z^2_n$ for the second). These detections were successful despite the large reduction in the X-ray flux (nearly one order of magnitude over a gap of $\sim4$\,years) with significances of $>5\sigma$ against the $\chi^2$ statistic for EF and $Z^2_n$ tests. The periodicity appears more pronounced when the source is fainter like $\zeta$ Tau (Fig. \ref{fig:HD43514_pulse}) and, indeed, the pulsed fractions in the two observations were measured to be 22\% and 34\%, respectively. The detection in the first observation is also far less significant than in the second and only identified in the EF search, although the significance of these detections considering the two observations is still greater than $4\sigma$. Unlike in $\zeta$ Tau, the pulsed fraction is slightly lower in both observations in the soft band (0.2--3.0\,keV) of 21\% and 31\%, compared to the hard band (3.0--12.0\,keV) of 29\% and 52\%. Newer data will soon show how this signal has evolved over an even longer baseline, in combination with the on-going dissipation of the decretion disc \citep{rauw_hd45314_2026}.

\subsubsection{V771 Sgr}
\label{subsec:discuss_v771sgr}

\begin{figure}
\centering
\includegraphics[width=0.9\linewidth]{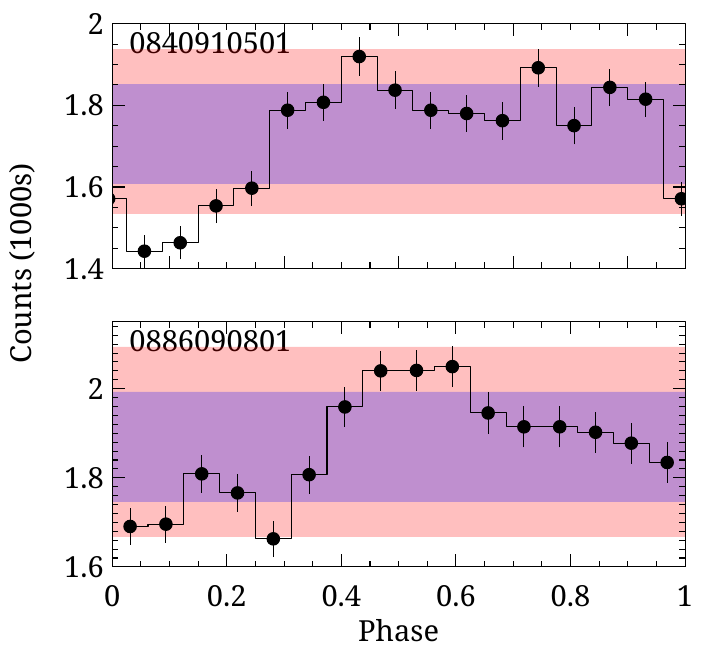}
  \caption{Same as Fig. \ref{fig:gcas_pulse1} but for V771 Sgr. The periodicity used for folding observation 0840910501 is 2.20\,ks while it is 2.06\,ks for 0886090801.}
     \label{fig:V771Sgr_pulse}
\vspace{-0.25cm}
\end{figure}

Previous periodicity searches identified signals at 2.1 and 4.2\,ks \citep{mondal_periodicity_2024,webbe_stonks_2026} in the second observation of this source. Here, GFA and $Z^2_n$ searches found a signal at $P=3.9\pm0.4$\,ks ($f=0.26\pm0.03$\,mHz) in the first observation (0840910801) and at $P=2.1\pm0.2$\,ks ($f=0.49\pm0.04$\,mHz) in the second one (0886090801) with significances $>5\sigma$ and $4\sigma$, respectively. We examined the first observation to determine whether there was a feature in the range $P=1.6-2.4$\,ks, which was significant at more than a 3-$\sigma$ level, but less significant than the strongest signal; furthermore, we identified a periodicity at $P=2.2\pm0.2$\,ks ($f=0.45\pm0.05$\,mHz) with a significance of $>5\sigma$, which is fully compatible with the other detection. The significance of these detections considering the two observations is greater than $4\sigma$. The pulsed fractions associated to that periodicity in the first and second observations are 14\% and 10\%, respectively (Fig. \ref{fig:V771Sgr_pulse}). Again, the detection appears more significant and the modulation is stronger when the source is fainter (i.e. in the first observation). The pulsed fraction is slightly lower in both observations in the soft band (0.2--3.0\,keV) of 14\% and 10\%, compared to the hard band (3.0--12.0\,keV) of 17\% and 13\%.

\subsubsection{$\pi$ Aqr}
\label{subsec:discuss_piaqr}

\begin{figure}
\centering
\includegraphics[width=0.9\linewidth]{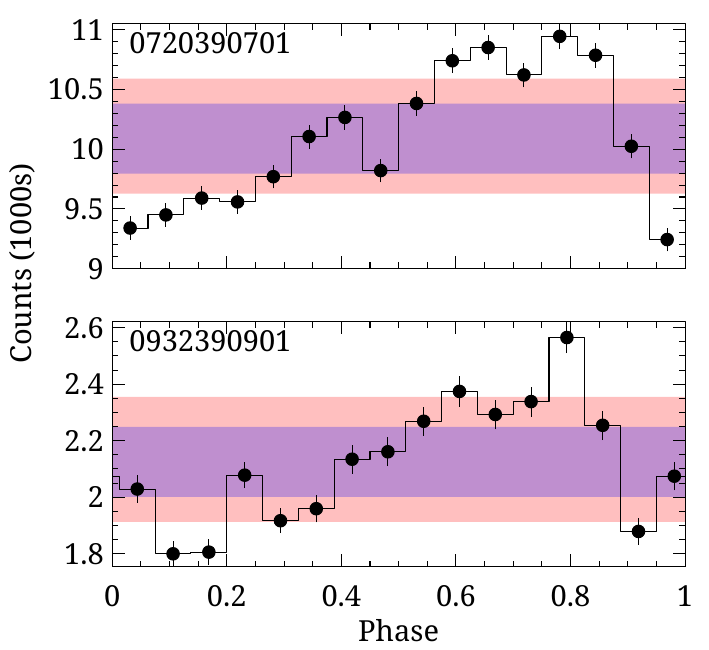}
  \caption{Same as Fig. \ref{fig:gcas_pulse1} but for $\pi$ Aqr. The periodicity used for folding observation 0720390701 is 1.79\,ks while it is 1.59\,ks for 0932390901.}
     \label{fig:PiAqr_pulse}
\vspace{-0.5cm}
\end{figure}

\citet{huenemoerder_chandra_2024} reported a 3.4\,ks periodicity for this source. We identified two different periodicities in the two \emph{XMM-Newton} observations ($P\sim10\pm1$\,ks and $P\sim1.6\pm0.3$\,ks with significances of $>5\sigma$ and $4\sigma$ respectively); however, we were not able to recover the previously identified periodicity in \emph{Chandra} observations \citep{huenemoerder_chandra_2024}. Notably, the observation lengths of these exposures were quite different ($T=54.4$\,ks and $T=10.0$\,ks, respectively), so we performed a targeted search in the range $f=0.56-0.71$\,mHz for the first, longer observation. It identified a significant signal at $P=1.8\pm0.1$\,ks ($f=0.56\pm0.02$\,mHz) with a significance $>5\sigma$, consistent within errors with the period detected in the second observation. The significance of these detections considering the two observations is greater than $4\sigma$. Moreover, several aliases of that peak can be seen, one of which lies near 0.64\,mHz (i.e. closer to the peak found in the second observation). Both values are close to half the period detected by \citet{huenemoerder_chandra_2024}. The profiles of these periodicities are remarkably similar, as shown in Fig. \ref{fig:PiAqr_pulse}. Their pulsed fractions amount to 8\% and 18\% in the first and second observations, respectively. As such, $\pi$ Aqr is the only one of these five candidates where the pulsed fraction increases with the source brightness. There is also no clear trend in the pulsed fraction with energy range in $\pi$ Aqr. In the first observation the pulsed fraction is higher in the hard band (11\% in 3.0-12.0\,keV) than the soft band (8\% in 0.2-3.0\,keV); however, this trend is reversed in the second observation, with pulsed fractions of 19\% in 0.2-3.0\,keV and 16\% in 3.0-12.0\,keV. The difference in pulsed fraction between the two bands is consistent within errors in the second observation, but not the first.

\subsection{Comparison with accreting white dwarfs}
\label{subsec:period_AWD_model}

For accreting magnetic WDs, we would expect a persistent periodic variability related to the rotation of the WD. This modulation of the light curve is related to the changing visibility of bright spots associated with the X-ray emission \citep[for reviews see ][]{kuulkers_x-rays_2006,mukai_x-ray_2017,webb_accreting_2023}. CVs also show periodic variability associated with the orbital period \citep[e.g.][]{kuulkers_x-rays_2006,parker_x-ray_2005,webb_accreting_2023}, but we did not consider this variability in the context of the present analysis as orbital phase-locked variations have not been detected thus far for $\gamma$\,Cas analogues \citep{naze_surprises_2019,rauw_x-ray_2022,naze_x-raying_2024} and the orbital periods of $\gamma$ Cas analogues are on timescales of 100s days.

The potential rotation periods of the WDs in $\gamma$ Cas analogues are unknown, as the impact of any accretion on the spin-up of the WD will be dependent on the geometry of the model and on the strength of the magnetic field \citep{frank_accretion_2002}. The rotation periods of isolated WDs can be significantly longer than is detectable in single X-ray observations, with periods of the order of tens of hours \citep{kawaler_rotation_2015,hermes_white_2017}; however, the periods of WDs in $\gamma$ Cas binaries could be significantly shorter if accretion has caused the WD to spin up \citep{kuulkers_x-rays_2006,mukai_x-ray_2017}. An analysis by \citet{naze_x-raying_2024} of $\zeta$ Tau identified that rotation periods of the order of 1\,ks could be achieved by accreting WDs \citep[see also][]{frank_accretion_2002} after 1\,Myr of activity. Therefore, the  periodicities identified for the five secure cases discussed above (1.2--5.4\,ks in $\gamma$ Cas, $\zeta$ Tau, HD45314, V771 Sgr, and $\pi$ Aqr) are indeed within the expected range.

The periodicities recovered in $\gamma$ Cas, $\zeta$ Tau, HD45314, V771 Sgr, and $\pi$ Aqr are also consistent with those seen in accreting WDs in close and wide binaries. The rotation periods of IPs are well-observed and many examples have been identified with $P_{\text{Rot}}$ of the order of 1\,ks \citep[e.g.][]{mukai_x-ray_2017,de_martino_hard_2020}. WDs in symbiotic star systems could provide a closer comparison to the WDs in the $\gamma$ Cas analogues, as these are broader binaries. However, there are fewer confirmed rotation periods available for the WDs in these systems. The only confirmed examples display comparable timescales to those measured here: CH Cyg (0.5\,ks \citealt{mikolajewski_symbiotic_1990}), Z And (1.7\,ks \citealt{sokoloski_discovery_1999}), and BF Cyg (6.5\,ks \citealt{formiggini_discovery_2009}). The spin period of Z And has been observed to decrease by $\sim0.08$\,ks over a period of 25\,years \citep{merc_accretion-induced_2024}. The rate of mass transfer and accretion in the $\gamma$ Cas analogues is expected to be greater than that seen in symbiotic stars, which would imply a faster period evolution in the former systems. We tentatively identified a marginal decreasing trend in the periodicities seen in $\gamma$ Cas for $P\sim1.7$\,ks (Table \ref{tab:gam-cas-per-grps}), and between periodicities observed in the older and newer observations of $\zeta$ Tau, HD45314, V771 Sgr, and $\pi$ Aqr. However, these differences remain below 2-$\sigma$ in all cases and require further confirmation.

The amplitudes of the periodic variability observed in $\gamma$ Cas, $\zeta$ Tau, HD45314, V771 Sgr, and $\pi$ Aqr are also consistent with those observed in CVs. The pulsed fractions recovered in the five sources discussed above lie in the $5-35\%$ range and are similar to those observed in a sample of hard X-ray selected CVs \citep{bernardini_broad-band_2017} in a comparable energy range (0.3-12\,keV). The profiles of the periodicities for each source vary across observations, as also observed in IPs \citep[e.g.][]{pekon_orbital_2011}. This may be due to changes in inclination or emitting region geometry with varying orbital phase, as expected and observed in IPs. The variation in the pulsed fraction across observations is likely due to changes in the visibility of the emitting regions and might be an indicator that the inner radius at which the accretion disc is truncated is changing, thereby causing an obscuration (partial or total) of the emitting regions on the surface of the WD. A time-domain and spectroscopic analysis of the periodicities in these sources dependent on orbital and rotation phase is required, however, this is beyond the current scope of this work.
\vspace{-0.4cm}

\section{Conclusions}
\label{sec:conc}

The X-ray emission from $\gamma$ Cas and its analogues was recently associated with an accreting WD, likely to be a magnetic one \citep{naze_xrism}. For such WDs, short-term periodicities are expected due to the WD rotation. In this work, we review the short-term variability in the \emph{XMM-Newton} and \emph{Chandra} observations of 26 $\gamma$ Cas analogues and 2 candidates. A total of 131 observations were analysed in a homogeneous way using various methods (i.e. Fourier analyses, light curve folding). 

The aperiodic 'flickering' variability was detected through increased Fourier amplitudes at low frequencies ($PSD$ or $A^2 \propto f^{-\alpha}$). The fitting of the power-law relationship was possible in more than a quarter of observations; however, Poisson noise prevented this fitting in the rest. The best-fit power-law indices were close to one, as usually observed in CVs and symbiotics. 

Significant features (greater than $3\sigma$) were detected in the periodicity searches in many cases, but only five systems appear to be secure detections of recurrent periodicities: $\gamma$ Cas, $\zeta$ Tau, HD45314, V771 Sgr, and $\pi$ Aqr. In these fives cases, the period is consistent within errors and the pulse profiles show some evolution between observations. In most other cases, further data are needed for confirmation of potential signals. In this context, we note that the data quality needs to be extremely high to make the detection possible. Finally, the detected periodicities lie within the 1.2--5.4\,ks interval, with pulsed fractions of 5--35\%. Again, such values are similar to what is observed for other accreting IP systems. In several cases, the periodic modulation appears stronger even when the target appears fainter. Detailed simulations of WD accretion in high-mass, long-period configurations are now required to improve our understanding of these properties.

%%%%%%%%%%%%%%%%%%%%%%%%%%%%%%%%%%%%%%%%%%%%%%%%%%%%%%%%%%%%%%
\begin{acknowledgements}
      Authors RW and NW acknowledge support from the CNES for this work. YN acknowledges support by the FNRS and Univ. of Li\`ege.
\end{acknowledgements}

\bibliographystyle{aa}
\bibliography{GamCasPeriod}

\begin{appendix}

\onecolumn
\section{Observations of $\gamma$ Cas analogues}
\label{app:obs-list}
We list the details of all observations of the $\gamma$ Cas analogues included in this analysis.

\begin{longtable}{lccccr}
\caption{Observations analysed in this work, listed chronologically for each source.}\\
\label{tab:all-obs} \\
\hline\hline
Name & Obs. & ObsID & Date & Exposure (ks) & Avg. Rate (cts s$^{-1}$) \\
\hline
\endfirsthead
\caption{continued.}\\
\hline
Name & Obs. & ObsID & Date & Exposure (ks) & Avg. Rate (cts s$^{-1}$) \\
\hline
\endhead
\hline
\endfoot
$\gamma$ Cas & Chandra & 1895 & 10-08-2001 & 52.5 & 0.381 \\
$\gamma$ Cas & XMM & 0201220101 & 05-02-2004 & 71.0 & 48.356 \\
$\gamma$ Cas & XMM & 0651670201 & 07-07-2010 & 17.9 & 41.632 \\
$\gamma$ Cas & XMM & 0651670301 & 24-07-2010 & 16.1 & 37.303 \\
$\gamma$ Cas & XMM & 0651670401 & 02-08-2010 & 17.9 & 55.198 \\
$\gamma$ Cas & XMM & 0651670501 & 20-08-2010 & 23.8 & 37.305 \\
$\gamma$ Cas & XMM & 0743600101 & 24-07-2014 & 34.0 & 53.054 \\
$\gamma$ Cas & XMM & 0840310101 & 04-01-2021 & 19.9 & 21.143 \\
$\gamma$ Cas & XMM & 0840310201 & 17-02-2021 & 18.5 & 42.155 \\
$\gamma$ Cas & XMM & 0840310301 & 25-07-2021 & 9.9 & 48.127 \\
$\gamma$ Cas & XMM & 0840310401 & 18-01-2022 & 17.6 & 36.859 \\
TYC 3681-695-1 & XMM & 0112200201 & 09-07-2002 & 9.0 & 0.053 \\
V782 Cas & XMM & 0102580801 & 11-03-2000 & 50.4 & 0.221\\
$\zeta$ Tau & Chandra & 26239 & 25-12-2021 & 11.6 & 0.255 \\
$\zeta$ Tau & XMM & 0920020301 & 20-03-2023 & 17.0 & 1.546 \\
$\zeta$ Tau & XMM & 0920020401 & 03-10-2023 & 22.0 & 1.665 \\
$\zeta$ Tau & Chandra & 30520 & 16-12-2024 & 10.1 & 0.205 \\
$\zeta$ Tau & Chandra & 30684 & 20-12-2024 & 10.1 & 0.156 \\
$\zeta$ Tau & Chandra & 30685 & 21-12-2024 & 19.3 & 0.180 \\
$\zeta$ Tau & Chandra & 30686 & 21-12-2024 & 10.0 & 0.178 \\
HD44458 & XMM & 0820310301 & 08-09-2018 & 15.9 & 0.513 \\
HD45314 & XMM & 0670080301 & 14-04-2012 & 26.2 & 0.224 \\
HD45314 & XMM & 0760220601 & 08-03-2016 & 31.0 & 0.039 \\
HD45995 & XMM & 0820310401 & 24-09-2018 & 13.0 & 0.324 \\
HD90563 & XMM & 0780070701 & 15-08-2016 & 8.0 & 0.025 \\
HD90563 & XMM & 0860650501 & 24-01-2021 & 49.0 & 0.053 \\
HD90563 & XMM & 0880000301 & 30-06-2021 & 17.0 & 0.064 \\
HD90563 & XMM & 0880000401 & 01-07-2021 & 31.3 & 0.018 \\
HD90563 & XMM & 0880000501 & 03-08-2021 & 22.0 & 0.079 \\
HD90563 & XMM & 0880000601 & 07-08-2021 & 24.4 & 0.101 \\
HD110432 & XMM & 0109480101 & 03-07-2002 & 53.0 & 5.612 \\
HD110432 & XMM & 0109480201 & 26-08-2002 & 48.7 & 4.805 \\
HD110432 & XMM & 0109480401 & 21-01-2003 & 48.0 & 4.875 \\
HD110432 & XMM & 0504730101 & 04-09-2007 & 80.7 & 2.474 \\
HD110432 & Chandra & 9947 & 13-11-2009 & 141.3 & 0.228 \\
HD110432 & XMM & 0840760201 & 21-07-2019 & 17.0 & 1.904 \\
HD119682 & XMM & 0087940201 & 28-08-2001 & 40.5 & 0.508 \\
HD119682 & Chandra & 4554 & 26-12-2004 & 15.1 & 0.105 \\
HD119682 & Chandra & 8929 & 17-12-2008 & 29.3 & 0.032 \\
HD119682 & Chandra & 10835 & 19-12-2008 & 30.0 & 0.024 \\
HD119682 & Chandra & 10834 & 20-12-2008 & 60.0 & 0.025 \\
HD119682 & Chandra & 10836 & 21-12-2008 & 30.1 & 0.028 \\
HD119682 & XMM & 0551000201 & 06-03-2009 & 57.3 & 0.342 \\
HD119682 & XMM & 0840310901 & 26-08-2019 & 14.3 & 0.301 \\
HD119682 & XMM & 0840311001 & 21-01-2020 & 25.3 & 0.332 \\
HD119682 & XMM & 0840311101 & 20-07-2020 & 14.5 & 0.095 \\
HD119682 & XMM & 0840310801 & 06-03-2021 & 14.2 & 0.219 \\
V767 Cen & XMM & 0402121801 & 25-01-2007 & 11.0 & 0.905 \\
V767 Cen & XMM & 0891800801 & 07-08-2021 & 12.0 & 0.973 \\
CQ Cir & XMM & 0840910401 & 20-09-2019 & 7.3 & 0.880 \\
HD157832 & XMM & 0551020101 & 05-09-2008 & 21.9 & 0.610 \\
HD157832 & XMM & 0810210301 & 03-04-2018 & 30.4 & 0.194 \\
HD157832 & Chandra & 27032 & 01-02-2023 & 13.8 & 0.033 \\
HD157832 & Chandra & 27689 & 03-02-2023 & 14.8 & 0.033 \\
HD157832 & Chandra & 27031 & 23-09-2023 & 20.0 & 0.027 \\
HD157832 & Chandra & 27033 & 29-01-2024 & 9.6 & 0.036 \\
HD157832 & Chandra & 29232 & 29-01-2024 & 11.7 & 0.033 \\
HD157832 & Chandra & 29233 & 29-01-2024 & 12.5 & 0.033 \\
HD157832 & Chandra & 29234 & 30-01-2024 & 9.4 & 0.029 \\
HD157832 & Chandra & 29374 & 20-04-2024 & 25.9 & 0.024 \\
HD157832 & Chandra & 26494 & 21-04-2024 & 15.1 & 0.028 \\
HD157832 & Chandra & 27034 & 12-06-2024 & 17.7 & 0.027 \\
HD157832 & Chandra & 29448 & 15-06-2024 & 18.1 & 0.030 \\
HD157832 & Chandra & 30463 & 03-10-2024 & 17.6 & 0.035 \\
HD157832 & Chandra & 27035 & 05-10-2024 & 14.0 & 0.029 \\
HD161103 & XMM & 0201200101 & 26-02-2004 & 17.8 & 0.401 \\
HD161103 & Chandra & 8647 & 16-05-2008 & 2.2 & 0.079 \\
HD161103 & XMM & 0691760101 & 08-09-2012 & 22.9 & 0.270 \\
V771 Sgr & XMM & 0840910501 & 22-03-2019 & 19.8 & 0.924 \\
V771 Sgr & XMM & 0886090801 & 02-10-2023 & 23.0 & 1.332 \\
HD316568 & XMM & 0206590201 & 05-09-2004 & 20.9 & 0.024 \\
HD316568 & XMM & 0402280101 & 10-09-2006 & 44.1 & 0.024 \\
2XMM J180816.6-191939 & XMM & 0024940201 & 22-03-2001 & 25.9 & 0.005 \\
GSC2 S300302371 & XMM & 0135741001 & 30-03-2001 & 8.5 & 0.035 \\
GSC2 S300302371 & XMM & 0840911001 & 05-04-2020 & 50.4 & 0.028 \\
SS397 & Chandra & 159 & 23-08-1999 & 17.7 & 0.015 \\
SS397 & Chandra & 1230 & 23-08-1999 & 15.8 & 0.011 \\
SS397 & Chandra & 162 & 27-08-1999 & 17.6 & 0.009 \\
SS397 & Chandra & 1233 & 05-11-1999 & 17.5 & 0.003 \\
SS397 & Chandra & 1433 & 15-11-1999 & 16.8 & 0.019 \\
SS397 & Chandra & 1717 & 23-05-2000 & 8.5 & 0.027 \\
SS397 & Chandra & 1718 & 23-05-2000 & 8.5 & 0.027 \\
SS397 & Chandra & 1722 & 23-05-2000 & 8.5 & 0.020 \\
SS397 & Chandra & 1723 & 23-05-2000 & 8.5 & 0.016 \\
SS397 & XMM & 0122700101 & 07-04-2000 & 35.3 & 0.115 \\
SS397 & XMM & 0122700201 & 09-04-2000 & 34.2 & 0.142 \\
SS397 & XMM & 0122700301 & 11-04-2000 & 34.2 & 0.095 \\
SS397 & XMM & 0122700401 & 15-04-2000 & 32.9 & 0.085 \\
SS397 & XMM & 0122700501 & 17-04-2000 & 32.6 & 0.148 \\
SS397 & XMM & 0122700801 & 01-04-2001 & 17.4 & 0.039 \\
SS397 & Chandra & 1554 & 21-07-2001 & 11.1 & 0.030 \\
SS397 & XMM & 0804250201 & 06-04-2017 & 42.0 & 0.017 \\
SS397 & XMM & 0890200101 & 10-03-2021 & 13.0 & 0.032 \\
SS397 & XMM & 0890200201 & 10-03-2021 & 13.9 & 0.008 \\
SS397 & XMM & 0890200301 & 10-03-2021 & 13.0 & 0.024 \\
SS397 & XMM & 0890200501 & 10-03-2021 & 13.0 & 0.032 \\
SS397 & XMM & 0890200601 & 10-03-2021 & 13.0 & 0.057 \\
SS397 & XMM & 0890200701 & 11-03-2021 & 13.0 & 0.037 \\
SS397 & XMM & 0890200801 & 11-03-2021 & 13.9 & 0.040 \\
SS397 & XMM & 0890200901 & 11-03-2021 & 21.8 & 0.013 \\
NGC 6649 9 & XMM & 0122700101 & 07-04-2000 & 35.3 & 0.134 \\
NGC 6649 9 & XMM & 0122700401 & 15-04-2000 & 32.9 & 0.140 \\
NGC 6649 9 & XMM & 0122700501 & 17-04-2000 & 32.6 & 0.058 \\
NGC 6649 9 & XMM & 0122701001 & 09-04-2001 & 19.1 & 0.146 \\
NGC 6649 9 & XMM & 0804250201 & 06-04-2017 & 42.0 & 0.028 \\
NGC 6649 9 & XMM & 0890200101 & 10-03-2021 & 13.0 & 0.037 \\
NGC 6649 9 & XMM & 0890200201 & 10-03-2021 & 13.9 & 0.046 \\
NGC 6649 9 & XMM & 0890200301 & 10-03-2021 & 13.0 & 0.090 \\
NGC 6649 9 & XMM & 0890200401 & 10-03-2021 & 13.0 & 0.128 \\
NGC 6649 9 & XMM & 0890200601 & 10-03-2021 & 13.0 & 0.076 \\
NGC 6649 9 & XMM & 0890200901 & 11-03-2021 & 21.8 & 0.102 \\
3XMM J190144.5+045914 & XMM & 0136030201 & 21-09-2003 & 20.2 & 0.036 \\
3XMM J190144.5+045914 & XMM & 0840910801 & 23-10-2019 & 39.9 & 0.025 \\
V558 Lyr & XMM & 0840200401 & 03-10-2019 & 9.0 & 0.573 \\
SAO 49725 & XMM & 0201200201 & 07-07-2010 & 11.5 & 0.218 \\
V2156 Cyg & XMM & 0840910601 & 30-11-2019 & 18.0 & 0.138 \\
$\pi$ Aqr & XMM & 0720390701 & 17-11-2013 & 54.5 & 2.254 \\
$\pi$ Aqr & Chandra & 26079 & 23-08-2022 & 10.1 & 0.153 \\
$\pi$ Aqr & Chandra & 27269 & 27-08-2022 & 10.1 & 0.140 \\
$\pi$ Aqr & Chandra & 26080 & 05-09-2022 & 30.1 & 0.122 \\
$\pi$ Aqr & Chandra & 26001 & 13-09-2022 & 20.1 & 0.096 \\
$\pi$ Aqr & Chandra & 27412 & 14-09-2022 & 22.1 & 0.129 \\
$\pi$ Aqr & Chandra & 27325 & 30-10-2022 & 9.9 & 0.155 \\
$\pi$ Aqr & XMM & 0932390901 & 26-11-2023 & 10.0 & 2.980 \\
V810 Cas & Chandra & 3453 & 15-04-2002 & 12.0 & 0.108 \\
\hline
Candidate &  &  &  &  &  \\ 
HD42054 & XMM & 0402121401 & 11-04-2007 & 15.7 & 1.048 \\
HD42054 & Chandra & 11021 & 21-06-2010 & 126.2 & 0.040 \\
HD42054 & Chandra & 12226 & 24-06-2010 & 26.0 & 0.036 \\
HD120678 & XMM & 0820310601 & 10-03-2019 & 50.0 & 0.039 \\
HD120678 & XMM & 0870590131 & 10-03-2020 & 5.3 & 0.147 \\
HD120678 & XMM & 0870590132 & 10-03-2020 & 6.6 & 0.209 \\
\end{longtable}
\tablefoot{Columns from left to right: source name, facility performing the observation (\emph{XMM-Newton} or \emph{Chandra}), observation ID, date the observation began, total exposure time of the observation (ks), and the observed count rate during the observation from \emph{XMM-Newton} (0.2-12\,keV) or \emph{Chandra} (0.3-10.0\,keV).}

\FloatBarrier
\vspace{2cm}

\section{Fitting to power spectra of GTIs}
\label{app:psds}
\begin{figure*}[h!]
\centering
\includegraphics[width=0.85\textwidth]{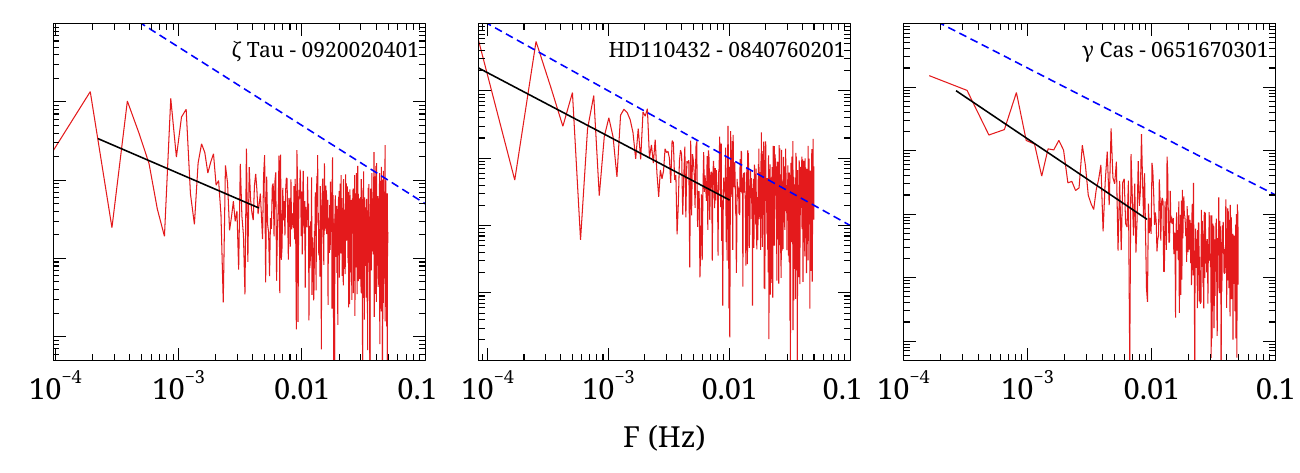}
  \caption{Examples of power-law fits to PSDs (red line) of light curves within GTIs in 3 observations that were sufficiently clear of Poisson noise to permit fitting a power-law relationship (black solid line). The blue dashed line indicates a power law of $P\propto f^{-1}$ for reference. In all cases the PSD appears visually to be consistent with a power law with an exponent of -1, subject to the observed noise and uncertainty over when the PSD is dominated by Poisson noise at higher frequencies. The three panels show: \emph{Left} - Observation of $\zeta$ Tau which is formally fit to a power law with an exponent of $-0.68\pm0.19$. \emph{Middle}: Observation of HD110432that is formally fit to a power law with an exponent of $-0.95\pm0.13$. \emph{Right}:Observation of $\gamma$ Casthat is formally fit to a power law with an exponent of $-1.31\pm0.19$.}
\label{fig:psd_fit_examples}
\end{figure*}

\clearpage

\end{appendix}
\end{document}